\documentclass[twocolumn,tighten]{aastex62}
\usepackage{array}
\usepackage{amsmath}
\usepackage{multirow}
\usepackage{graphicx}
\usepackage{placeins}
\usepackage[utf8]{inputenc}
\usepackage[caption=false]{subfig}
\usepackage[percent]{overpic}
\usepackage{color}

\begin{document}

\def\fesc{\textit{f$_{esc}$}}
\def\hst{\textit{HST}}

\newcommand\setrow[1]{\gdef\rowmac{#1}#1\ignorespaces}

\title{Candidate $z \sim 2.5$ Lyman Continuum Sources in the GOODS Fields}

\author[0000-0002-1706-7370]{L.~H.~Jones}
\affiliation{Department of Astronomy, University of Wisconsin-Madison,
475 N. Charter Street, Madison, WI 53706, USA}

\author[0000-0002-3306-1606]{A.~J.~Barger}
\affiliation{Department of Astronomy, University of Wisconsin-Madison,
475 N. Charter Street, Madison, WI 53706, USA}
\affiliation{Department of Physics and Astronomy, University of Hawaii,
2505 Correa Road, Honolulu, HI 96822, USA}
\affiliation{Institute for Astronomy, University of Hawaii, 2680 Woodlawn Drive,
Honolulu, HI 96822, USA}

\author[0000-0002-6319-1575]{L.~L.~Cowie}
\affiliation{Institute for Astronomy, University of Hawaii,
2680 Woodlawn Drive, Honolulu, HI 96822, USA}


\begin{abstract}
We use the wealth of deep archival optical spectroscopy on the GOODS-South field from Keck, the VLT, and other facilities to select candidate high-redshift Lyman continuum (LyC) leakers in the \textit{Hubble} Deep UV Legacy Survey (HDUV) dataset. We select sources at $2.35 < z < 3.05$, where the \emph{HST}/WFC3 F275W filter probes only the redshifted LyC. We find five moderately F275W-bright sources (four detected at $\gtrsim3\sigma$ significance) in this redshift range. However, two of these show evidence in their optical spectra for contamination by foreground galaxies along the line-of-sight. We then perform an F275W error-weighted sum of the fluxes of all 129 galaxies at $2.35 < z < 3.05$ in both the GOODS-N and GOODS-S HDUV areas to estimate the total ionizing flux. The result is dominated by just five candidate F275W-bright LyC sources. Lastly, we examine the contributions to the metagalactic ionizing background, finding that, at the sensitivity of the HDUV F275W data and allowing for the effects of LyC transmission in the intergalactic medium (IGM), star-forming galaxies can match the UV flux required to maintain an ionized IGM at $z \sim 2.5$.
\end{abstract}


\section{Introduction}
\label{intro}

For the past few decades, one of the most active areas of research in observational astronomy has been the identification of sources that contribute to the metagalactic ionizing background. This has been particularly important for building a cohesive picture of reionization --- an important epoch in the history of the Universe in which the first galaxies formed and the bulk of hydrogen in the intergalactic medium (IGM) transitioned from neutral to ionized (e.g., \citealt{bou06,bou12,bou15,ouch09,rob15}). Massive stars and quasars/active galactic nuclei (AGN) both produce ionizing photons, though their relative importance to the global ionizing background appears to evolve over time. Most evidence currently favors a scenario in which dwarf star-forming galaxies (SFGs) are the primary driver of hydrogen reionization \citep{ric00,bou06,font07,font14,rob10,rob15,jap17}, while AGN contributions to the ionizing background are small until $z \sim$ 2 -- 3 \citep{barg03,bolt05,cbt09,cri16,smith20}. However, because these scenarios typically rely on an extrapolation of observed galaxy counts to faint absolute magnitudes unobtainable even in the deepest \textit{HST} imaging, some authors have argued that quasars/AGN could produce a non-negligible or even dominant fraction of UV photons during the era of reionization (e.g., \citealt{font12,mad15}). Of course, arguments for or against this alternate scenario depend critically on the assumed number density of AGN at high redshift, which is still a subject of significant debate (see, e.g., \citealt{gia19,cowie20,graz20} and \citealt{shin20} for some recent analyses).

Another major uncertainty in determining the relative importance of SFGs and AGN to reionization is constraining the escape fraction $f_{esc}$, defined as the fraction of all Lyman continuum (LyC, rest frame $\lambda$ $<$ 912 \AA{}) photons that manage to escape their galaxy of origin to interact with the IGM. Most theoretical and semi-analytical models of reionization require an average $f_{esc}$ for SFGs of about 10\% or greater, if SFGs are to be the primary driver of reionization (e.g., \citealt{bolt07,van12a,feng16,price16,kimm17}; see, however, \citealt{fauch08} and \citealt{matt17}), though at the highest redshifts, $f_{esc}$ remains largely unconstrained by observations. For sources at $z \gtrsim$ 4, the rising density of intervening neutral hydrogen absorption systems leads to much lower transmissivity of the IGM to LyC photons \citep{mad95,song04,in14}. This effectively prohibits direct detections of LyC emission (and hence estimates of $f_{esc}$) along most sightlines at redshifts where such measurements are most needed, though some exceptions have been reported in recent years (e.g., the $z = 4$ source \textit{Ion3} in \citealt{van18} and the $z = 3.8$ source \textit{Ion1} in \citealt{ji20}). Thus, observations focused on analogous objects at slightly lower redshifts are often used to constrain the ionization history of the Universe.

Literature reports of individual or stacked LyC detections suggest that most galaxies have small values of $f_{esc}$, at most $\sim1-3\%$ (e.g., \citealt{lei95,stei01,grim09,cow10,leit13,rut16}), and that the typical escape fraction increases with redshift (e.g., \citealt{mit13,font14,fai16,kai16,jap17}). The latter observation is based on a growing number of LyC detections at $z \sim 2-3$ (e.g., \citealt{van10b,van12b,van18,most15,graz16,shap16,jones18,stei18,laces,riv19,saha20}), though it is now well known that direct LyC searches can be affected by contamination from foreground galaxies (e.g., the projections from \citealt{van10a} or the reexamination of sources first reported in \citealt{shap06} by \citealt{siana15}). Stacking analyses also tend to give a relatively weak average LyC signal at these redshifts (e.g., \citealt{siana10,graz17,rut17,marchi18,nai18,laces}; see, however, \citealt{stei18}). 

However, some studies have suggested that particular subgroups of galaxies are more likely to have significant LyC escape. For example, moderately large Ly$\alpha$ equivalent width in emission (e.g.,  \citealt{mich17,marchi18,stei18,laces}) or multiply-peaked Ly$\alpha$ line profiles \citep{verh17, van20} appear to be signposts of nonzero $f_{esc}$; see, however, the confirmed LyC leaker \textit{Ion1}, which shows Ly$\alpha$ in absorption only \citep{ji20}. A high flux ratio of [OIII]$\lambda\lambda4959,5007$ to [OII]$\lambda\lambda3727,3729$ (O32) has also recently been proposed as an optical marker of LyC escape (e.g., \citealt{nakajima14,izo16a,izo16b,izo18,stei18,laces,tang19}). However, this connection remains tenuous, with several studies reporting that large O32 by itself is insufficient to guarantee significant LyC escape (e.g., \citealt{reddy16b, izo17, rut17, nai18, barr20}). 

Legacy fields like the GOODS-North and South \citep{goods04} tend to be especially attractive targets for LyC leaker searches, due to the abundance of deep multiwavelength imaging and thorough spectroscopic coverage. In particular, the advent of the \textit{Hubble} Deep UV (HDUV) Legacy Survey \citep{oesch18} has now made it possible to perform direct photometric searches for LyC leakers at high redshift. Using the HDUV's deep \hst{}/WFC3 imaging in the F275W and F336W bands, \citet{nai17} identified six candidate LyC sources in the GOODS fields at $z \sim 2$, all with $f_{esc} \gtrsim$ 13\%. However, at the redshifts probed by \citet{nai17}, the Lyman break sits squarely in the middle of the F275W bandpass. Since both ionizing and non-ionizing photons contribute to each object's F275W photometry, a number of modeling assumptions are needed to sift out the contribution of just the LyC to the F275W flux. 

However, if one were to push to slightly higher redshifts, say $z \gtrsim 2.4$, as we did in the GOODS-N in \citet{jones18}, then the F275W filter becomes sensitive to LyC photons \textit{only}. In that work, we identified a raw total of six sources with spectroscopic redshifts $z \gtrsim 2.4$ that remained bright in F275W. However, four of these were then shown via optical spectroscopy to be line-of-sight blends of low- and high-redshift galaxies. 

In this paper, we turn our attention to the GOODS-S, using the wealth of optical/IR spectroscopy in this field in concert with deep F275W imaging from the HDUV survey, to search for individual candidate LyC-leaking galaxies at $z >$ 2.35. We place constraints 
on the mean LyC signal at this redshift using an averaging analysis and on the contributions to the overall ionizing luminosity density from SFGs. In Section \ref{data}, we describe the data we used to select and characterize potential high-redshift LyC leakers, including optical/IR redshift catalogs and spectra, and UV, X-ray, and optical imaging. We present our search for individual candidate LyC leakers in Section \ref{indiv_results} and discuss the properties of our candidate sources, along with evidence for or against contamination by foreground galaxies for each. In Section~\ref{averaging}, we perform an averaging 
analysis of all $z = 2.35 - 3.05$ sources with F275W coverage in the GOODS-N, in the GOODS-S, and in the two fields combined, from which we measure their total F275W contributions. In Section~\ref{stacking}, we stack the F275W and F336W images for each GOODS field and for the two
fields combined---first for $z = 2.35 - 3.05$, and then for three redshift bins for the combined images to look for differences that might arise due to ``dilution'' from Lyman absorption along the line-of-sight, which increases with redshift. In Section \ref{flux}, we estimate the metagalactic ionizing background at $2.35<z<3.05$ and compare to the flux required to maintain an ionized IGM at these redshifts.
Finally, we summarize our findings in Section~\ref{summary}.

We assume $\Omega_{M}$ = 0.3, $\Omega_{\Lambda}$ = 0.7, and $H_{0}$ = 70~km~s$^{-1}$~Mpc$^{-1}$ throughout this work. All magnitudes are given in the AB system; magnitude zero-points for the HDUV F275W and F336W observations are given in \citet{oesch18}.

\section{Data}
\label{data}
\subsection{F275W Imaging}
\label{datahduv}
The HDUV survey (GO13872; \citealt{oesch18}) is a 132-orbit WFC3 imaging program centered on the GOODS-North and South fields. Designed to capitalize on existing WFC3/UVIS imaging from the CANDELS \citep{grog11,koek11} and UVUDF \citep{tep13,rafel15} programs, the HDUV survey imaged both of the GOODS fields in the F275W and F336W filters around or within the existing CANDELS and UVUDF footprints. 

We focused on the HDUV GOODS-N dataset in \citet{jones18}.  Here we focus on the HDUV GOODS-S dataset, which includes one science image per filter and corresponding maps of the per-pixel rms, all drizzled to a 60 mas pixel resolution. The final HDUV images incorporate data from the UVUDF survey, which substantially deepens this region of the field. We do not include data from the region of the CDF-S field observed by the Early Release Science (ERS; \citealt{ers}). This area is almost a magnitude shallower in F275W than the regions that we use, and it contains significantly more artifacts; thus, we are unable to construct a high-significance $<26$ magnitude sample for that region.

When measuring our F275W fluxes (see Section~\ref{uvsample}), we smoothed the F275W and F336W images using the \texttt{scipy.ndimage.gaussian\_filter} function with a standard deviation of 2 to minimize the impact of residual noise on our measurements, though as we will show, these residuals still contribute significantly to our flux uncertainties.

\subsection{Optical/NIR Spectroscopy}
\label{zspec}
Secure spectroscopic redshifts are required for reliable identification of LyC-leaking candidates, as photometric redshift estimates and by-eye inspections are frequently contaminated by galaxies along the line-of-sight (e.g., \citealt{van10a,siana15,jones18}). We therefore narrow our search only to those galaxies with robust spectroscopic redshifts between 2.35 and 3.05. At the lower redshift bound, the Lyman limit sits at 3055.2 \AA, at which point the total throughput of the F275W filter has declined to 2.3\% on the red side. Adopting a slightly stricter redshift minimum of $z = 2.37$, at which point the Lyman limit sits at 3073.4 for a total F275W throughput of 1\%, does not remove any of our candidates, nor does it substantially change the results of our stacking analyses in Section \ref{averaging}.

Similarly, the upper redshift bound relates to the wavelength at which the F336W filter probes only the redshifted LyC. Although we considered F275W-bright candidates with redshifts up to $z=3.55$ in the GOODS-N in \citet{jones18}, in this work we use $z = 3.05$ as the upper redshift bound in both the GOODS-N and GOODS-S subsamples for consistency. In principle, ionizing sources at $z > 3.05$ could also be detected in F336W using the methods described below, though the rapidly increasing opacity of the IGM to LyC photons at $z = 3 - 4$ would make such a search somewhat more difficult (e.g., \citealt{in14}). Since the LyC is redshifted fully into the F275W bandpass at $2.35 < z < 3.05$, the deep and relatively wide F275W coverage provided by the HDUV survey enables us to search for potential sources of ionizing radiation at high redshift.

We consulted spectroscopic catalogs from the literature (\citealt{popesso09,balestra10,cowie12,kurk13,morris15,bacon17,inami17,vandels,barg19,musewide}), along with unpublished spectra that we have obtained with Keck/DEIMOS,
to identify sources in the GOODS-S in the redshift range of interest. We limited ourselves to objects with redshifts marked as high-quality/confidence by the respective authors. For example, we only included spectroscopic redshifts from the MUSE-Deep and MUSE-Wide surveys that are flagged as confidence level 3, which represent unambiguous, multiple-line detections (for MUSE-Deep; \citealt{bacon17} and \citealt{inami17}) or identifications with ``very high certainty" (for MUSE-Wide; \citealt{musewide}). 

Our final list of candidates (see Section \ref{uvsample}) has spectroscopic redshifts from the MUSE-Deep program and from the catalogs of \citet{balestra10} and \citet{kurk13}.

\subsection{X-ray Imaging}
\label{xrays}
To identify X-ray counterparts to our F275W sample (see Section \ref{uvsample}) and to weed out probable AGN, we used X-ray data and catalogs from the 7 Ms \textit{Chandra X-ray Observatory} exposure of the \textit{Chandra} Deep Field-South \citep{luo17}, which reaches a limiting flux of $f_{0.5-2 \mathrm{ keV}} \approx 6.4 \times 10^{-18}$~erg~cm$^{-2}$~s$^{-1}$ near the image center. There are 95 F275W sources with X-ray counterparts within a 1\farcs5 search radius. We computed the rest-frame 2 -- 8 keV luminosities, $L_{X}$, of these counterparts from the 0.5 -- 2 keV fluxes with an assumed $\Gamma = 1.8$ and no absorption correction using
\begin{equation}
L_X = 4\pi d^2_L f_{0.5-2 \mathrm{keV}}\bigg(\frac{1+z}{4}\bigg)^{\Gamma-2}\ \mathrm{erg\ s^{-1}.}
\end{equation}
%
%
Of the sources with X-ray counterparts, four have $L_X > 10 ^{42}$ erg s$^{-1}$, which classifies them as AGN. One of these is a BALQSO.  However, none of our candidate high-redshift LyC-leaking sample (defined below) have X-ray counterparts.

\section{Search for Individual $\lowercase{z} \sim 3$ Candidate LyC Leakers in the GOODS-S}
\label{indiv_results}

\begin{figure}[bht]
\includegraphics[width=3.33in]{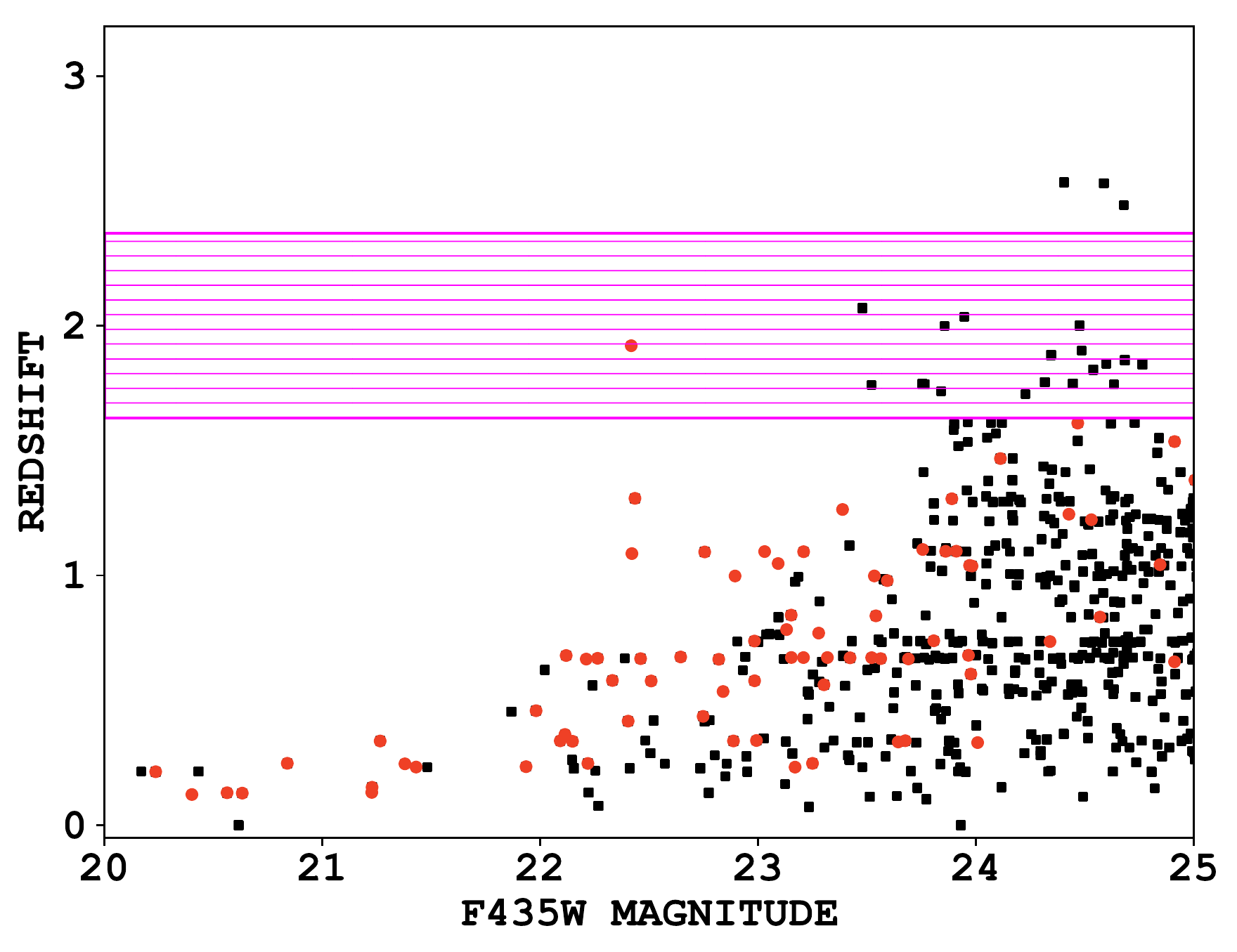}
\includegraphics[width=3.33in]{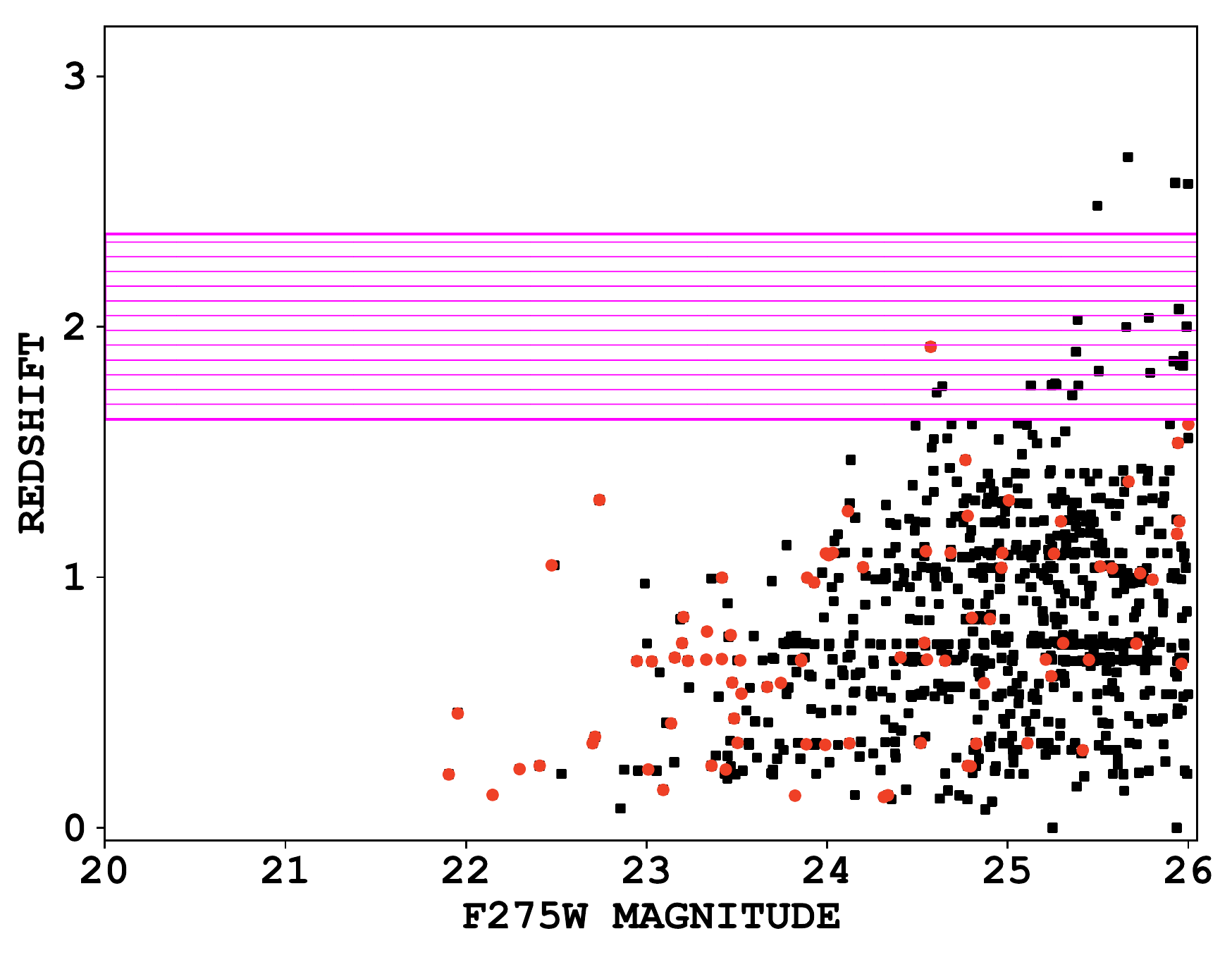}
\caption{\emph{Top:} Spectroscopic redshift vs. \textit{HST}/ACS F435W magnitude for all galaxies in the 43.5~arcmin$^2$ footprint of the HDUV GOODS-S with F850LP $<$ 26, which defines our parent sample. 
\emph{Bottom:} Spectroscopic redshift vs. \textit{HST}/WFC3 F275W magnitude for our primary UV sample of 1115 sources in the GOODS-S with F275W magnitudes $<$ 26 and total errors fainter than 27.19, corresponding to a $3\sigma$ detection. Red circles mark sources with an X-ray counterpart in the \textit{Chandra} Deep Field South 7 Ms catalog. The purple shaded region shows the redshift interval where the F275W bandpass straddles the Lyman limit. The four $> 3\sigma$ LyC-leaking candidates lie above this region.
}
\label{uvz}
\end{figure}

\citet{jones18} describes our search for LyC-leaking candidates in the HDUV GOODS-N field. In the GOODS-S, we begin with all F850LP $<$ 26 galaxies that are covered by the 43.5~arcmin$^2$ footprint of the HDUV observations. This area lies entirely within the GOODS-S observations of \citet{goods04} obtained with \hst{}'s Advanced Camera for Surveys (ACS). At $z \sim$ 2.5, the ACS F850LP filter probes the rest-frame FUV at $\sim$2500 \AA, providing a reasonable selection of star-forming galaxies at these redshifts. In the top panel of Figure~\ref{uvz}, we plot redshift versus F435W magnitude for this area. Spectroscopic identifications are 97\% complete below F435W = 24, 85\% from F435W = 24$-$24.5, and 63\% from F435W = 24.5$-$25.

\subsection{F275W Measurements in the GOODS-S}
\label{uvsample}

Using the Python Source Extraction and Photometry (SEP) library\footnote{https://github.com/kbarbary/sep} \citep{sep}, which is based on the SExtractor program of \citet{sextractor}, we measured F275W fluxes in $2\arcsec$ and $4\arcsec$ diameter apertures at the position of each F850LP $<$ 26 source with UV coverage. We also estimated the local median background at each source position in a $4 - 8\arcsec$ diameter annulus and subtracted this from the aperture fluxes. Comparing background-subtracted small- and large-aperture magnitudes (for extended objects with $21 <$ F275W $< 25$, to avoid saturated or low S/N sources) reveals a median offset of -0.18 mag, which we added back to the $2\arcsec$ magnitudes as an aperture correction.

We initially measured magnitude errors in $2\arcsec$ apertures from the associated rms noise files, with the same -0.18 mag aperture correction applied. However, this resulted in the selection of objects that, though they are detected nominally at moderate ($>3\sigma$) significance in F275W, appeared to be pure noise upon visual inspection. 

To estimate the effect of residual noise in the science maps on our flux measurements, we performed the same aperture photometry methods described above at $\sim$10,000 random positions across the HDUV field in both the F275W science and rms images, with the same smoothing applied for consistency with our source flux measurements. If we were to use only the rms image to determine uncertainties, we would underestimate the true noise by a factor of 2. This translates to an offset of -0.376 mag, which we added to our initial magnitude errors. 

Our primary UV sample consists of 1115 sources with F275W magnitudes brighter than 26 and total errors fainter than 27.19, corresponding to a $>3\sigma$ detection. In the bottom panel of Figure~\ref{uvz}, we show redshift versus F275W magnitude for this sample. Spectroscopic identifications are nearly fully complete to F275W = 24.5 (with only two sources out of 248 missing or unidentified), 92\% complete from F275W $= 24.5 - 25$, and 57\% complete from F275W $= 25 - 26$. The rapid drop-off in spectroscopic completeness at fainter UV magnitudes is a major motivation for our F275W = 26 cutoff, even though one might expect the number of candidate $z \sim 2.5-3$ LyC leakers to increase at very faint UV magnitudes. We have four LyC-leaking candidates in the redshift range $2.35 < z < 3.05$ from this selection.

\begin{figure}[t]
\includegraphics[width=3.33in]{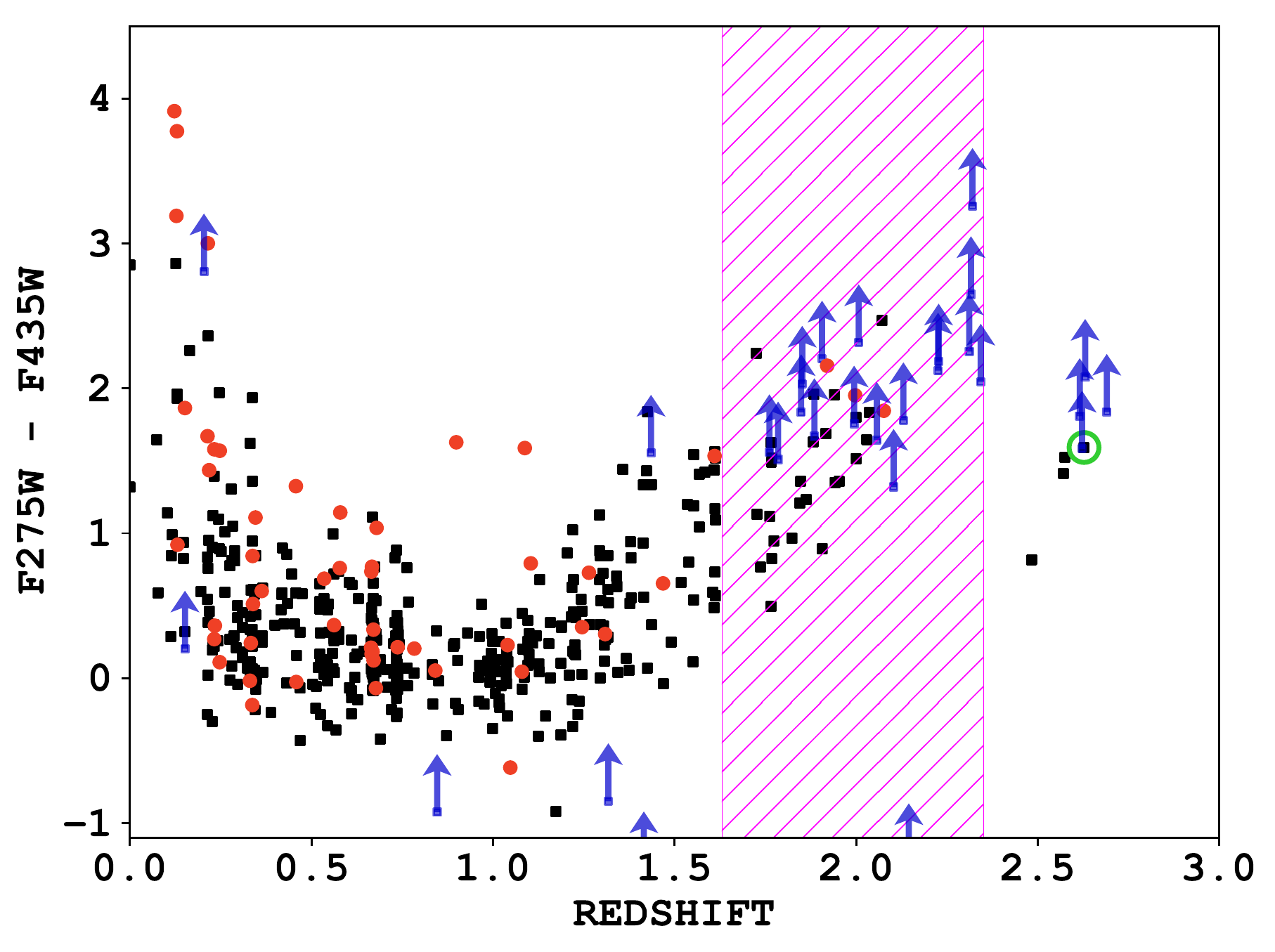}
\caption{F275W $-$ F435W color vs. spectroscopic redshift for sources with relatively flat rest-frame UV continua. Red circles mark sources with an X-ray counterpart in the \textit{Chandra} Deep Field South 7 Ms catalog. Sources with lower limits on F275W $-$ F435W are plotted at their 2$\sigma$ values with blue arrows. The open green circle marks the single $2.35 < z < 3.05$ source detected at $>2\sigma$ significance in F275W that was not already selected with our $>3\sigma$ significance criterion. The purple hatched region shows the redshift interval where the F275W bandpass straddles the Lyman limit.}
\label{color_z}
\end{figure}

Requiring a $>3\sigma$~detection in F275W at the depth of the HDUV imaging will necessarily miss UV-fainter sources that otherwise may merit further consideration as LyC-leaking candidates. We therefore used the following color selection as a secondary way of selecting potential LyC leakers. We required (1) F435W  $<$ 25; (2) F606W  $-$ F850LP  $<$ 1 mag; and (3) F275W signal-to-noise ratio (SNR) $>$ 2. The first condition ensures that this color-selected sample remains fairly spectroscopically complete, while the other conditions select sources with at least marginal F275W detections and relatively flat rest-frame, non-ionizing UV continua. As shown in Figure \ref{color_z}, this selection yields four candidates above our minimum redshift threshold, three of which were already selected directly using our $>3\sigma$ criterion. The remaining source (enclosed in the green circle in Figure \ref{color_z}) is detected at 2.6$\sigma$ in F275W. 

In summary, we have identified five candidate high-redshift LyC leakers in the redshift range $2.35 < z < 3.05$ in the GOODS-S, four of which are detected at $>3\sigma$ significance in F275W. With a total of 46 sources with UV coverage and secure redshifts in the redshift range, this leads to a $\sim 10.9\%$ raw success rate in identifying candidate LyC-leaking galaxies. However, after removing two candidates contaminated by foreground galaxies (see Section~\ref{lyccandidates}), this success rate drops to $\sim6.5\%$.

\begin{figure*}[ht]
\centering
\begin{tabular}{p{1.8in}p{1.8in}p{1.8in}p{1.8in}}

    \includegraphics[width=0.8\linewidth,height=0.8\linewidth]{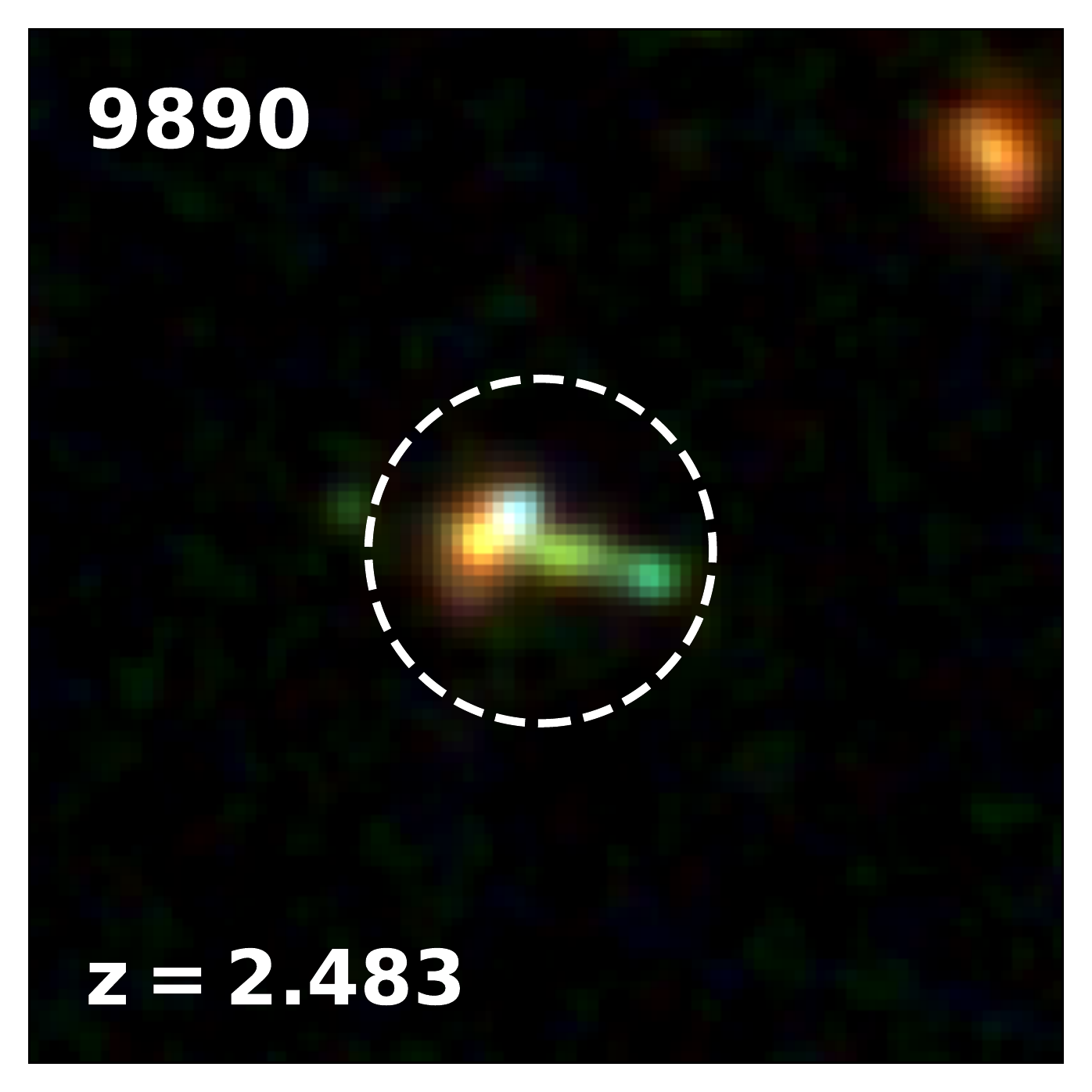}  & \hspace{-1cm}
    \includegraphics[width=0.8\linewidth,height=0.8\linewidth]{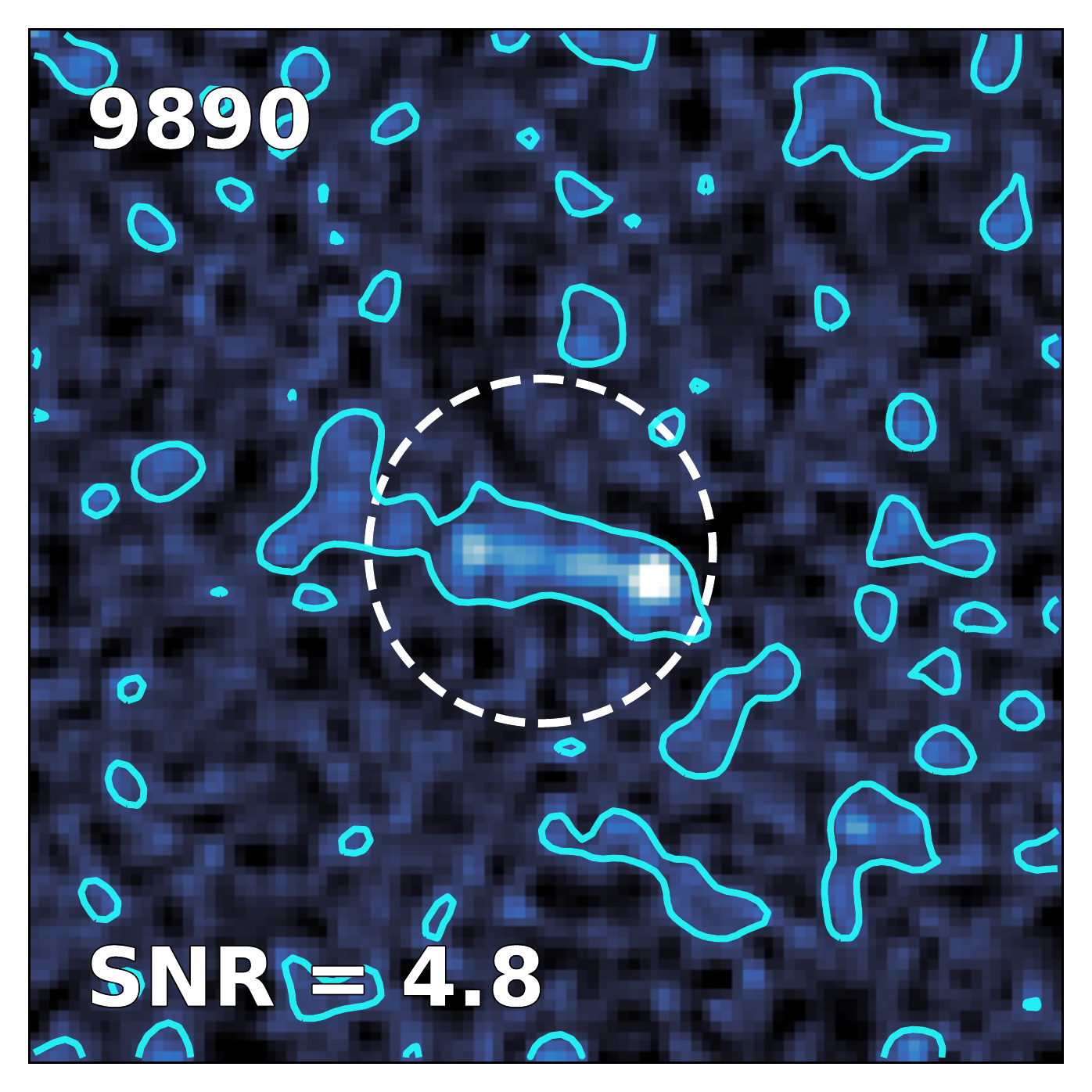}  & 
    \includegraphics[width=0.8\linewidth,height=0.8\linewidth]{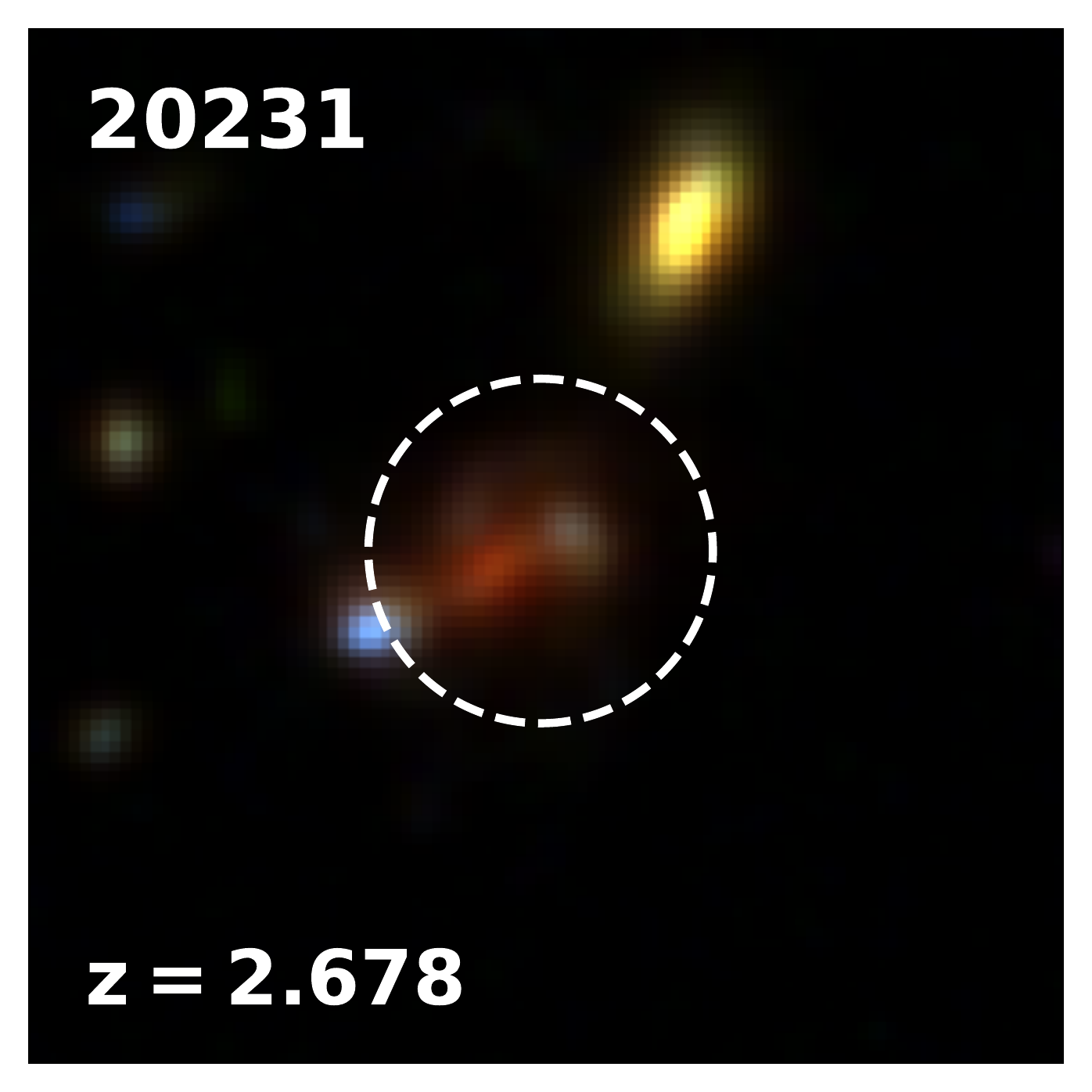} & \hspace{-1cm}
    \includegraphics[width=0.8\linewidth,height=0.8\linewidth]{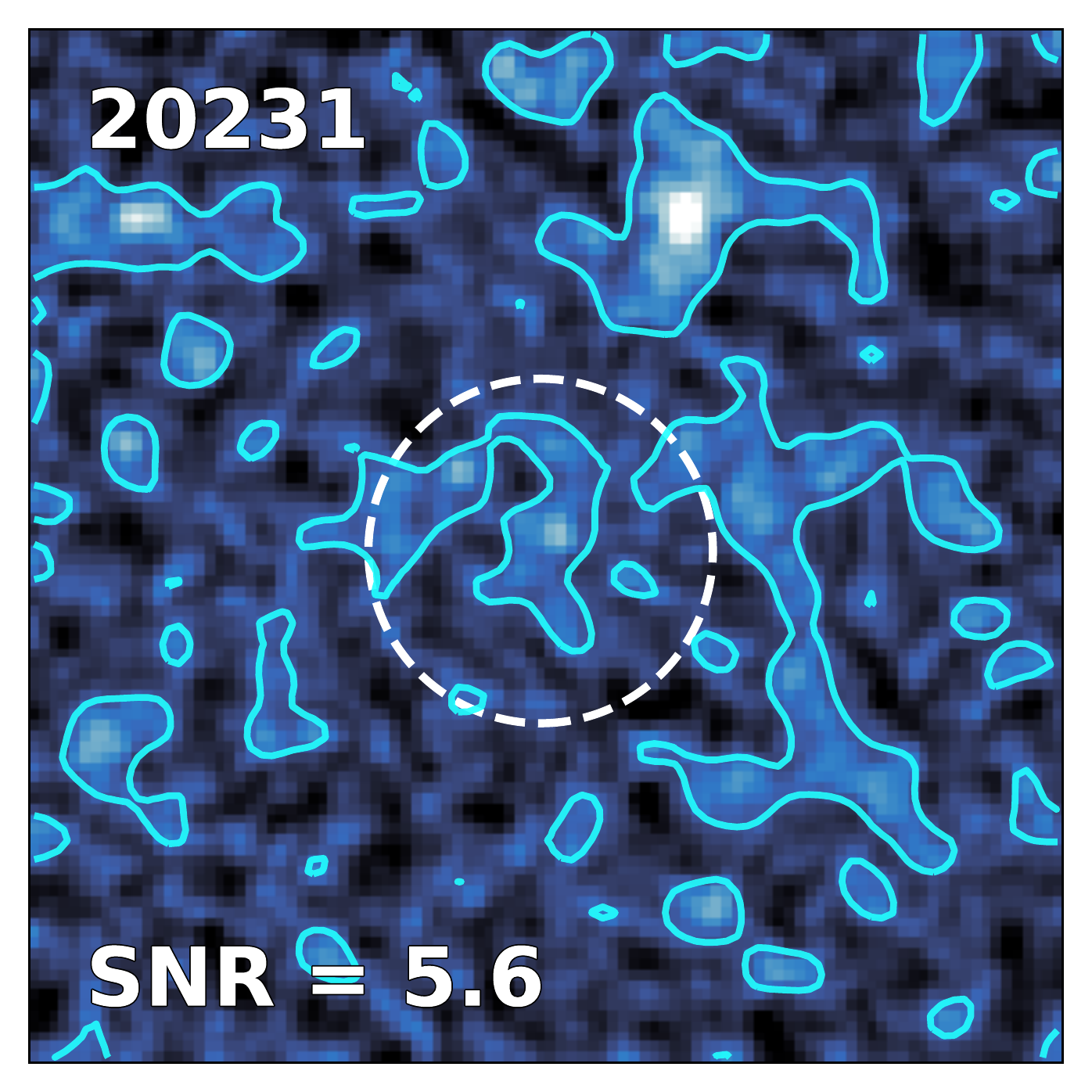}
     \\

  \includegraphics[width=0.8\linewidth,height=0.8\linewidth]{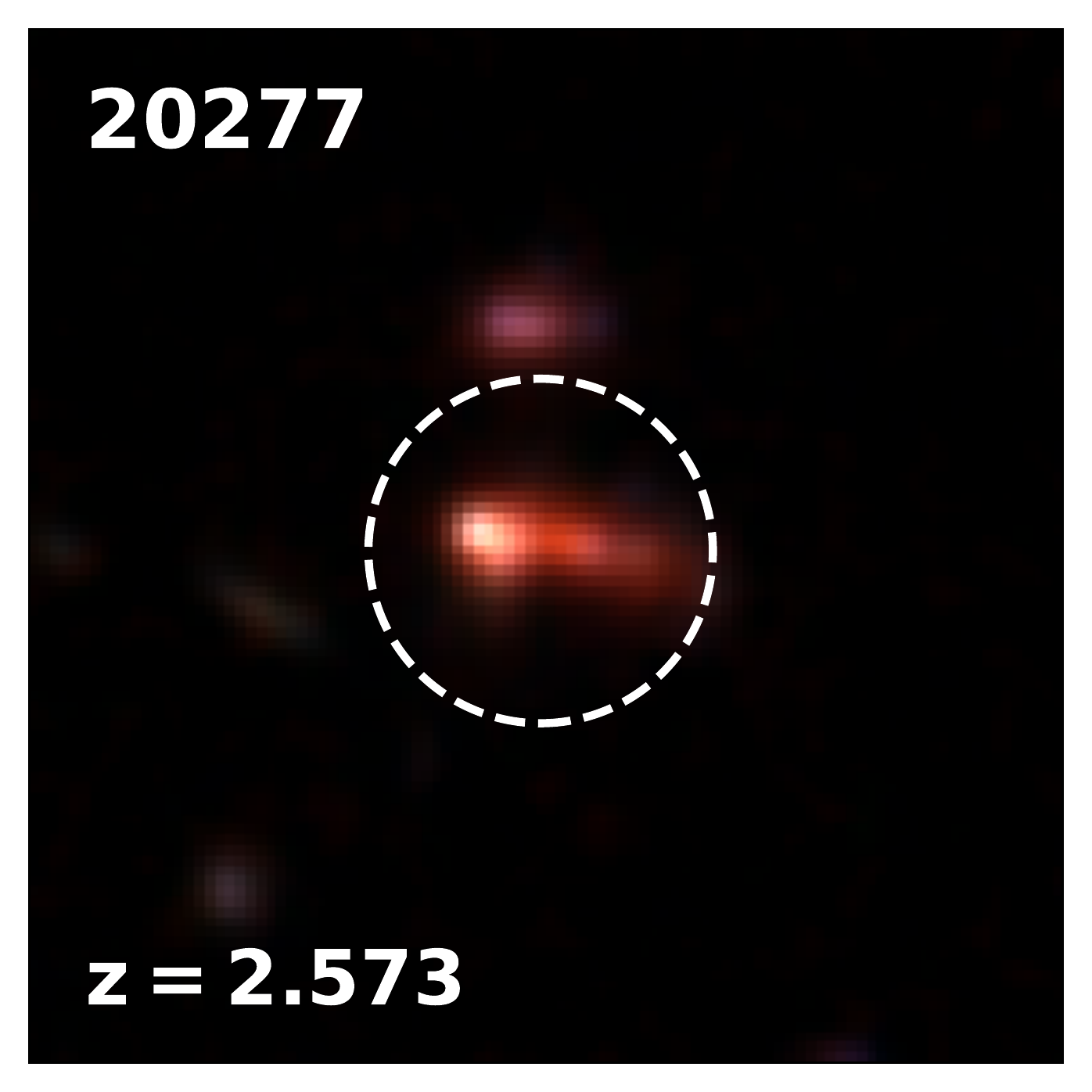} & \hspace{-1cm}
  \includegraphics[width=0.8\linewidth,height=0.8\linewidth]{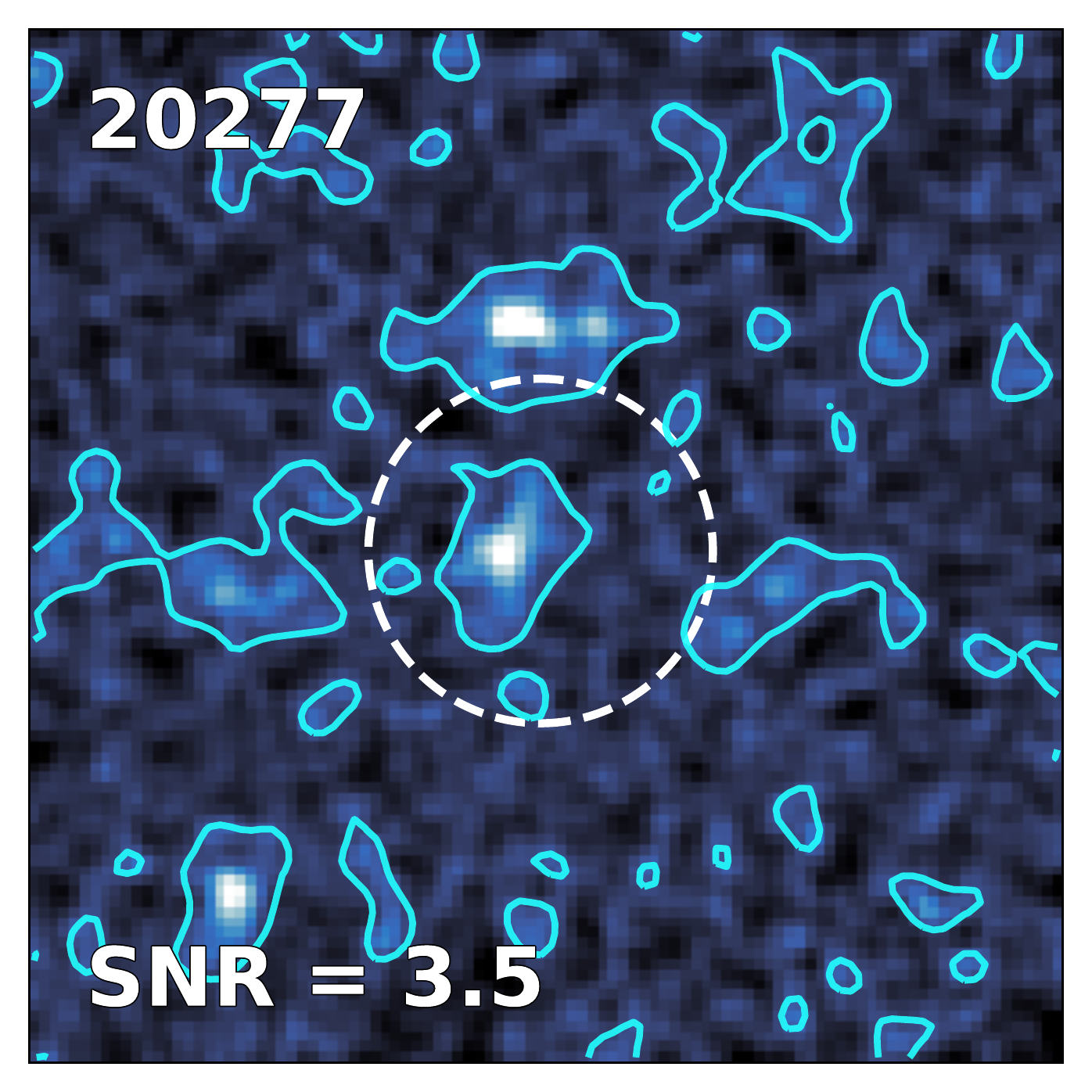} & 
  \includegraphics[width=0.8\linewidth,height=0.8\linewidth]{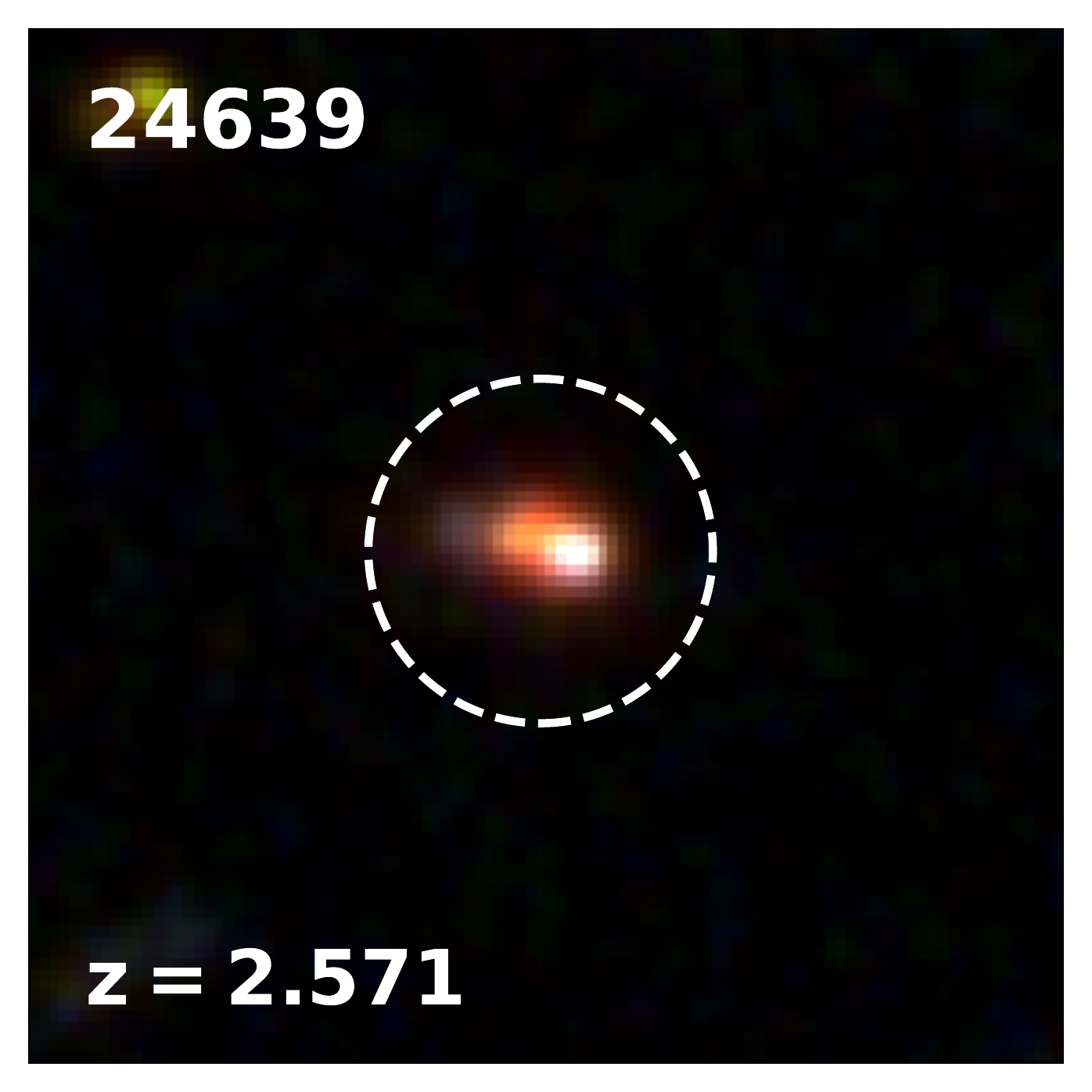}  & \hspace{-1cm}
  \includegraphics[width=0.8\linewidth,height=0.8\linewidth]{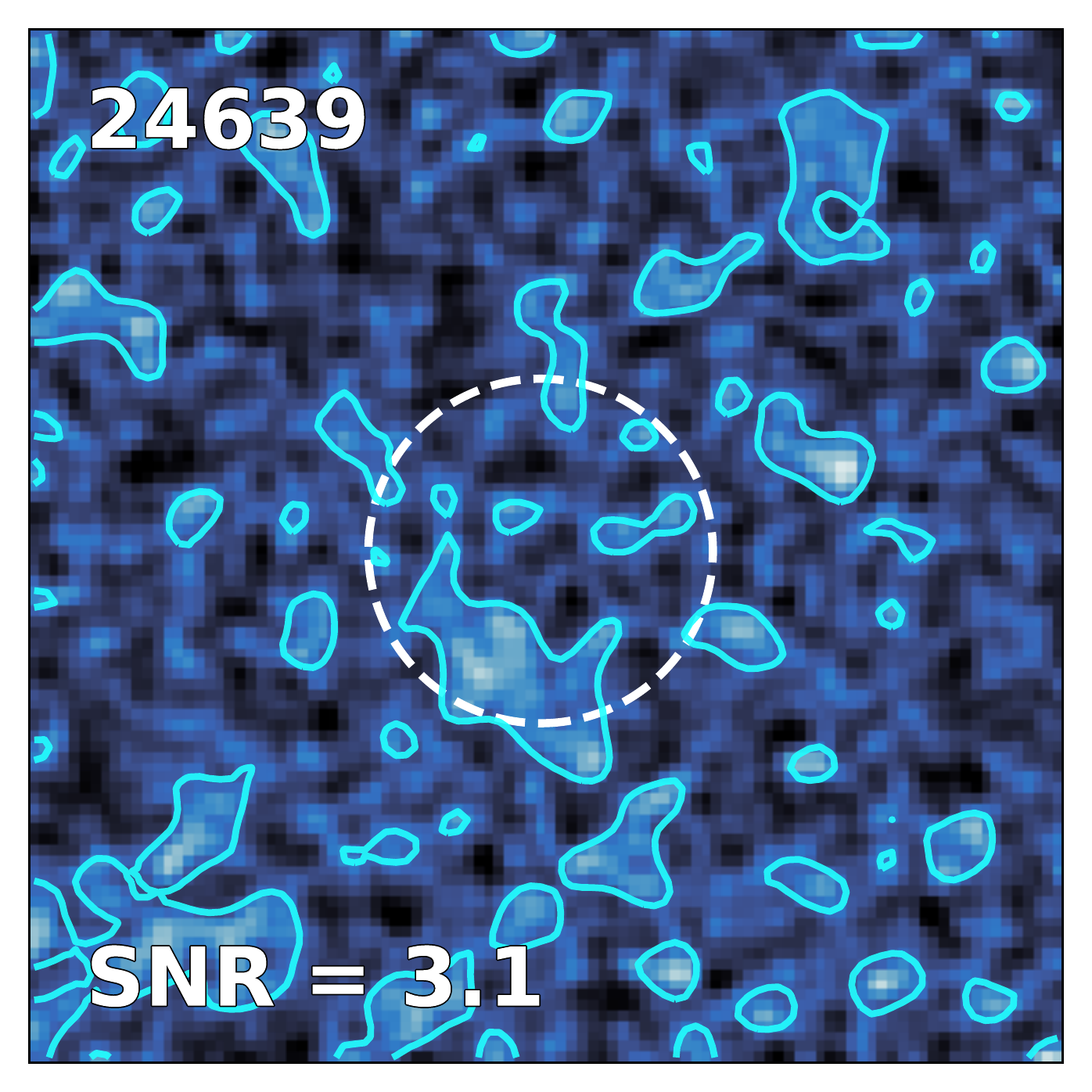} \\
  
  \hspace{0.49\textwidth}
  \includegraphics[width=0.8\linewidth,height=0.8\linewidth]{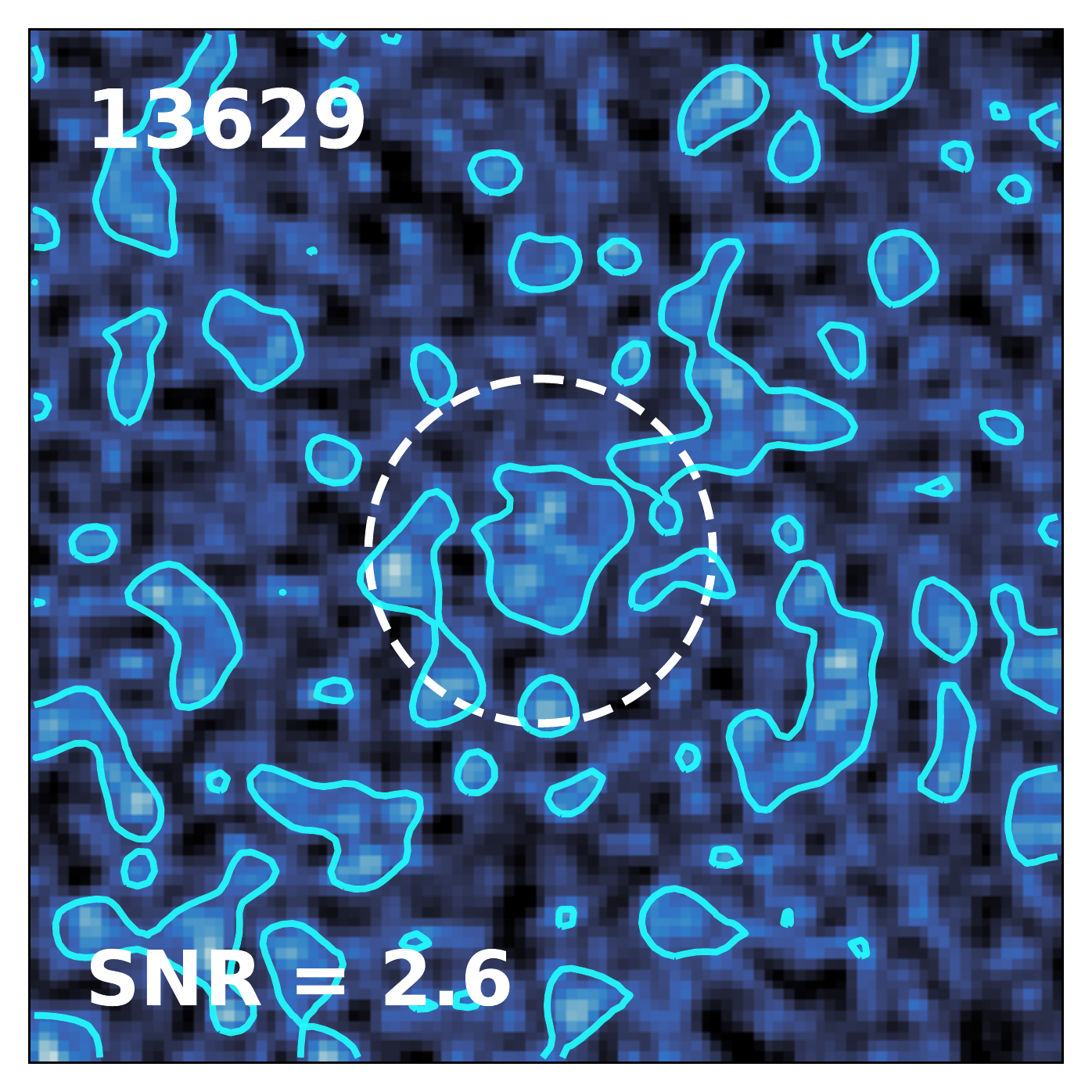} &
  \includegraphics[width=0.8\linewidth,height=0.8\linewidth]{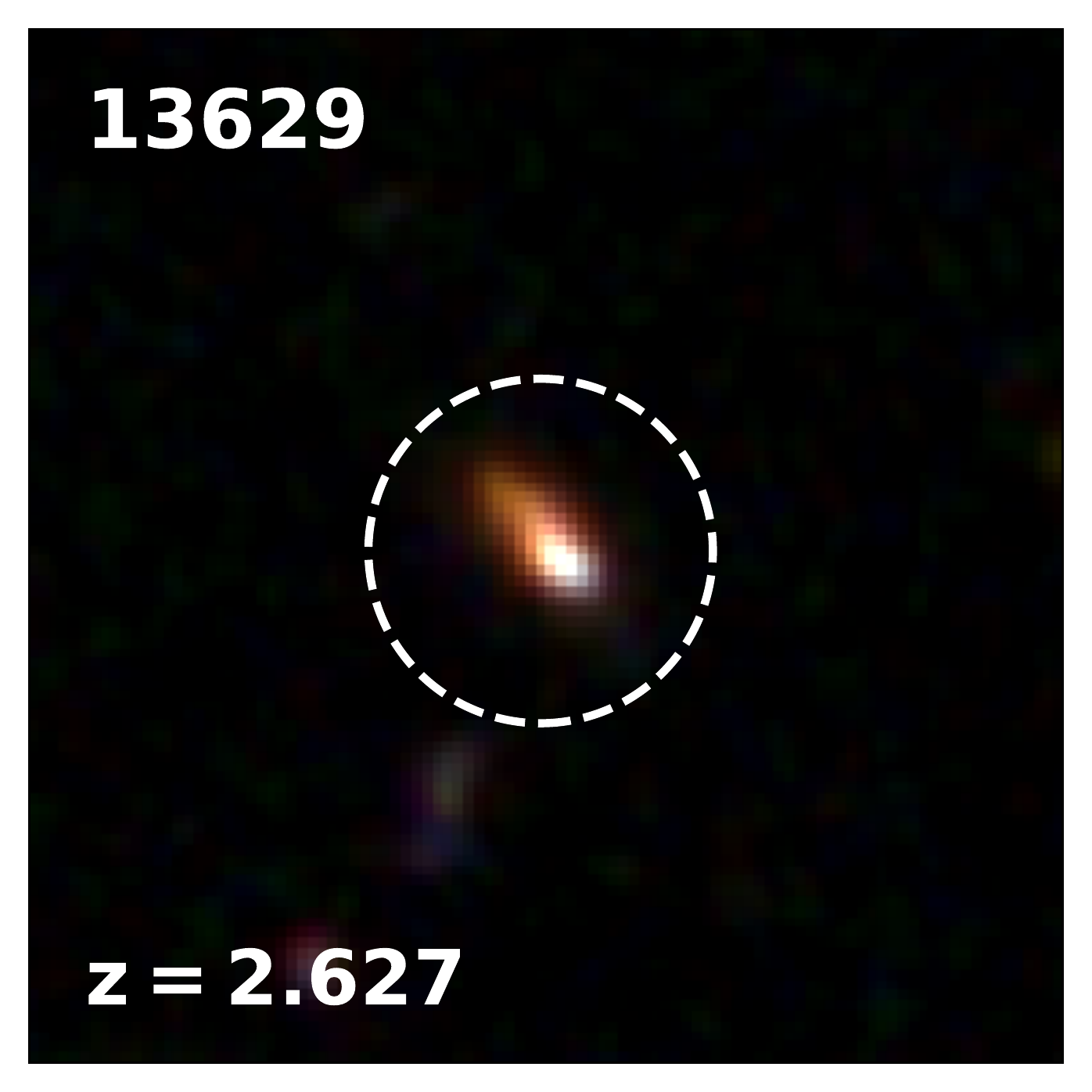} \\

\end{tabular}
\caption{
Three-color images (\emph{left}; red = F160W, green = F850LP, and blue = F435W) and F275W thumbnails (\emph{right}) of our five candidate LyC leakers. Cyan contours show regions where the F275W per-pixel SNR exceeds 5$\sigma$, and the white dashed circle shows the $2''$  diameter aperture used in measuring the magnitudes. Images are $6''$ on a side; North is up and East is to the left. For each source, the redshift (F275W SNR) is given at the bottom of the left (right) image. Sources 9890 and 20277 were found to be contaminated by low-redshift interlopers.
}
\label{thumbnails}
\end{figure*}  

\subsection{Five Candidate LyC Leakers in the GOODS-S}
\label{lyccandidates}
In Table \ref{tab:sources}, we list the basic properties of our five candidate LyC leakers in the GOODS-S. This includes an ID number, identical to those given in Table 3 of \citet{guo13}; decimal coordinates; a spectroscopic redshift from ground-based observations;
the F275W magnitude and error (including the -0.18 mag aperture correction and -0.376 mag noise correction); F435W magnitude; and F606W -- F850LP color.
In Figure \ref{thumbnails}, we show both a three-color thumbnail (left; red = F160W, green = F850LP, blue = F435W) and the F275W thumbnail (right) of each source.

{\setlength{\extrarowheight}{3pt}
\begin{table*}[t]
    \normalsize
    \centering
    \caption{\\\textsc{Summary of Candidate LyC Leakers}}
    \begin{tabular}{c c c c c c c}
    \hline
    \hline
    ID$^{a}$ & R.A. & Dec. & $z_{spec}^{b}$ 
    & F275W (Error)  & F435W  & F606W -- F850LP  \\
    \hline
    \setrow{\bfseries}9890$^{1}$ & 53.096661 & -27.772339 & 2.483 & 
    25.50 (27.21) & 24.68 & 0.57  \\
    20231$^{2}$ & 53.157532 & -27.798981 & 2.678 
    & 25.67 (27.54) & 25.49 & 0.33 \\
    \setrow{\bfseries}20277$^{3}$ & 53.157845 & -27.814756 & 2.573 
    & 25.93 (27.30) & 24.40 & 0.39  \\
    24639$^{3}$ & 53.188301 & -27.829344 & 2.571 
    & 26.00 (27.22) & 24.58 & 0.21  \\
    \hline
    13629$^{1}$ & 53.118439 & -27.805323 & 2.627 
    & 26.18 (27.24) & 24.59 & 0.16  \\
    \hline
    \end{tabular}
\label{tab:sources}
\tablecomments{$^a$Object IDs are the same as in Table 3 of \citet{guo13}. Sources with ID numbers in bold were later confirmed to be contaminated by foreground galaxies upon inspecting their spectra. \\
$^b$Spectroscopic redshifts from (1) GMASS \citep{kurk13}, (2) MUSE-Deep \citep{bacon17,inami17},\\ or (3) \citealt{balestra10}. \\
}
\end{table*}
}

Although we selected our candidates to have high-quality spectroscopic redshifts from $2.35 < z < 3.05$, the optical spectra for two sources (objects 9890 and 20277 in Table \ref{tab:sources}) show evidence of an intermediate redshift galaxy along the line-of-sight to the high-redshift galaxy. In source 9890, this is betrayed by the presence of an [OIII]$\lambda\lambda4959,5007$/H$\beta$ complex at $z = 0.784$. In source 20277, we see a strong [OII]$\lambda\lambda3727,3729$ doublet at $z = 1.094$. At these intermediate redshifts, the F275W filter probes the rest-frame FUV at $\sim 1300-1500$ \AA, well above the Lyman limit. The UV fluxes for these two sources are almost certainly dominated by the foreground galaxies and hence cannot be used to measure the ionizing flux output at $z \sim 2.5$. 

We note that these contaminated galaxies, particularly object 9890, reinforce the need for deep optical and NIR spectroscopy to confirm the nature of any proposed LyC leakers. In the case of object 9890, the photometric redshift without HDUV data ($z_{phot} = 0.698$; \citealt{3dhst}) is more consistent with the spectroscopic redshift of the foreground, low-redshift galaxy. Yet once the HDUV data are included, the photometric redshift better matches the spectroscopic redshift of the background galaxy ($z_{phot} = 2.36$; \citealt{oesch18}), meaning a selection based only on photometric redshifts would falsely label this source as a LyC leaker.

We now briefly discuss our remaining three candidate LyC leakers in the GOODS-S and the foreground contamination rate in our overall candidate sample. First, three distinct components are seen in the three-color thumbnail for source 20231 (upper-right of Figure \ref{thumbnails}): a generally red galaxy with two clumps, both of which are at almost-identical spectroscopic redshifts (R. Bacon, private communication), though it is unclear whether they constitute a single galaxy or an interacting pair. In any case, the F275W flux is not coincident with the blue clump to the southeast, but instead with the relatively red knot of emission near the image center.

Second, in source 24639 (middle-right of Figure \ref{thumbnails}), the F275W emission is offset from the position of the non-ionizing flux from this galaxy by $\sim0\farcs8$, which corresponds to about 6.4 proper kpc at $z = 2.571$. Though regions of LyC escape may not always be fully coincident with the bulk of the stellar emission, the lack of non-ionizing flux at the position of the F275W emission in this source is unusual. We do not remove this object from the list of LyC candidates, but in subsequent sections we note the effects of excluding it from our calculations.

Third, there is a similar off-center clump of F275W emission to the east of source 13629. However, unlike the case for source 24639, object 13629 also shows hints of F275W emission coincident with the position of the non-ionizing flux from this galaxy.

To estimate the expected level of foreground contamination, we ran Monte Carlo simulations to randomize the positions of all galaxies in the HDUV GOODS-S field, regardless of redshift identification or F275W magnitude. This assumes a spatially uniform foreground F275W population with a surface density of around 486,900 deg$^{-2}$ (the number of galaxies within the HDUV GOODS-S footprint divided by its 43.5 arcmin$^{2}$ area), similar to the assumptions in \citet{van10a}. For a sample size of 5, at least one galaxy is contaminated in 27\% of our 500 simulations, with a mean of 0.3 $\pm$ 0.6 contamination events. This is consistent with expectations from a simple binomial probability distribution, in which the probability of any single contamination event is proportional to the source surface density and the aperture size \citep{nestor13}. For the surface density given above and an aperture of radius 1\arcsec, an average of 0.48 $\pm$ 0.66 contamination events are expected in a sample of size 5.

Finally, we must also be concerned about the limitations of our sample selection in that we are only using sources with known spectroscopic redshifts. As we have argued above, it is dangerous to use photometric redshifts, which may be biased against continuum leakers, and where sources with incorrect high redshifts may also result in incorrect high fluxes. However, we have inspected the 185 SFGs that would be placed in the $z$ = 2.35 -- 3.05 redshift range by the \citet{straat16} catalog of photometric redshifts in the field and do not have spectroscopic redshifts. These have only two $3\sigma$ F275W detections and no significant total signal ($-0.25 \pm 0.58~ \mu$Jy).

Although the apparent offsets between the ionizing and non-ionizing radiation in some of these sources (particularly object 24639) are peculiar, we retain all of these objects as candidate LyC leakers. We emphasize that an absence of evidence \textit{for} contamination by foreground galaxies does not equate to positive evidence that such foreground contaminants are absent. For this reason and out of an abundance of caution, we assert that our three remaining sources in the GOODS-S should only be considered LyC-leaking \textit{candidates}.

\section{Mean ionizing emission from an averaging analysis using both GOODS fields}
\label{averaging}

In addition to searches for individual LyC leakers, the depth and breadth of UV coverage offered by the HDUV data enable a robust measurement of the ``typical" ionizing flux output of $z \sim 2.5$ galaxies via averaging. 

\begin{figure}[t]
\includegraphics[width=3.33in,trim={0.5cm 1cm 1cm 0},clip]{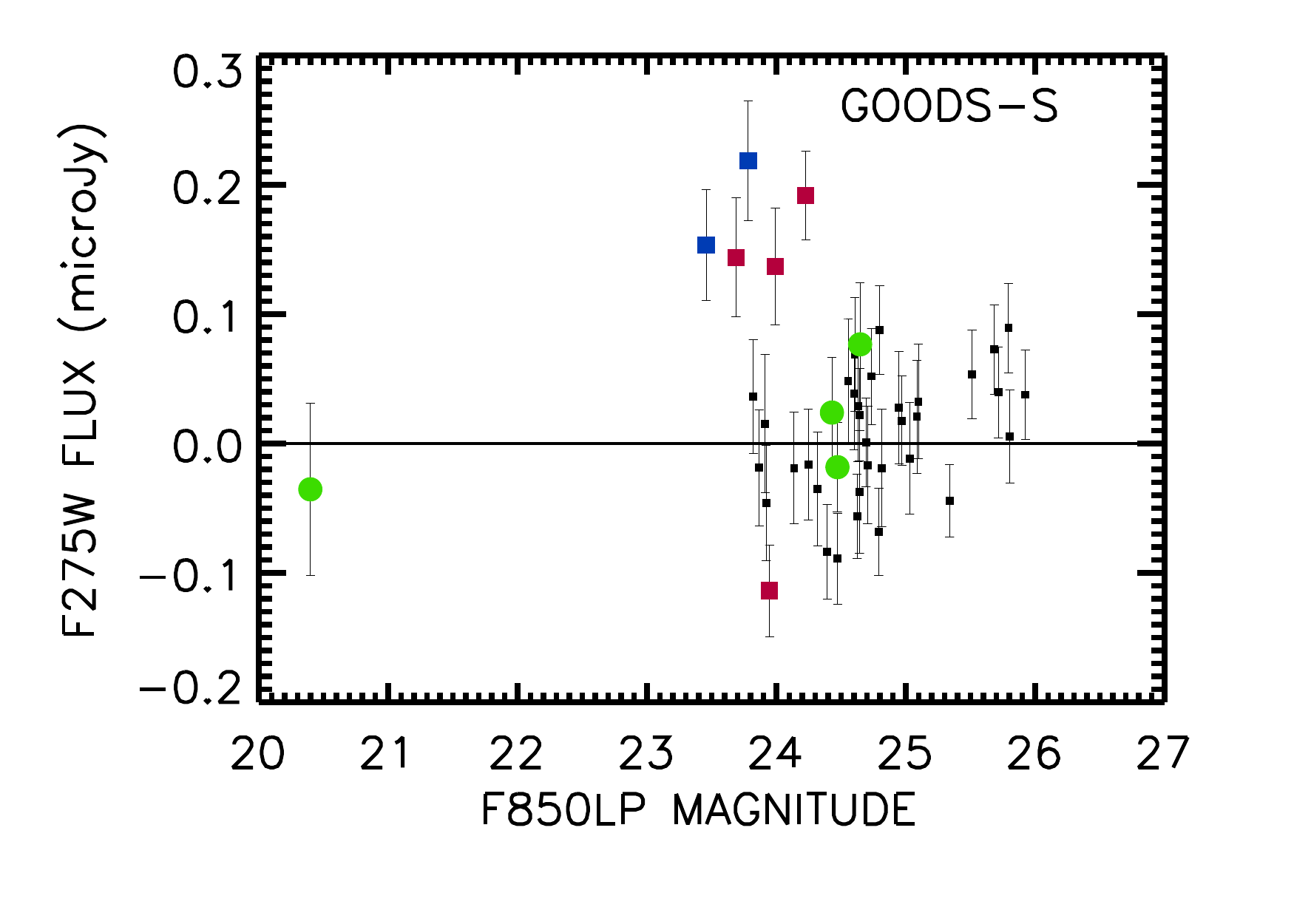}
\includegraphics[width=3.33in,trim={0.5cm 1cm 1cm 0},clip]{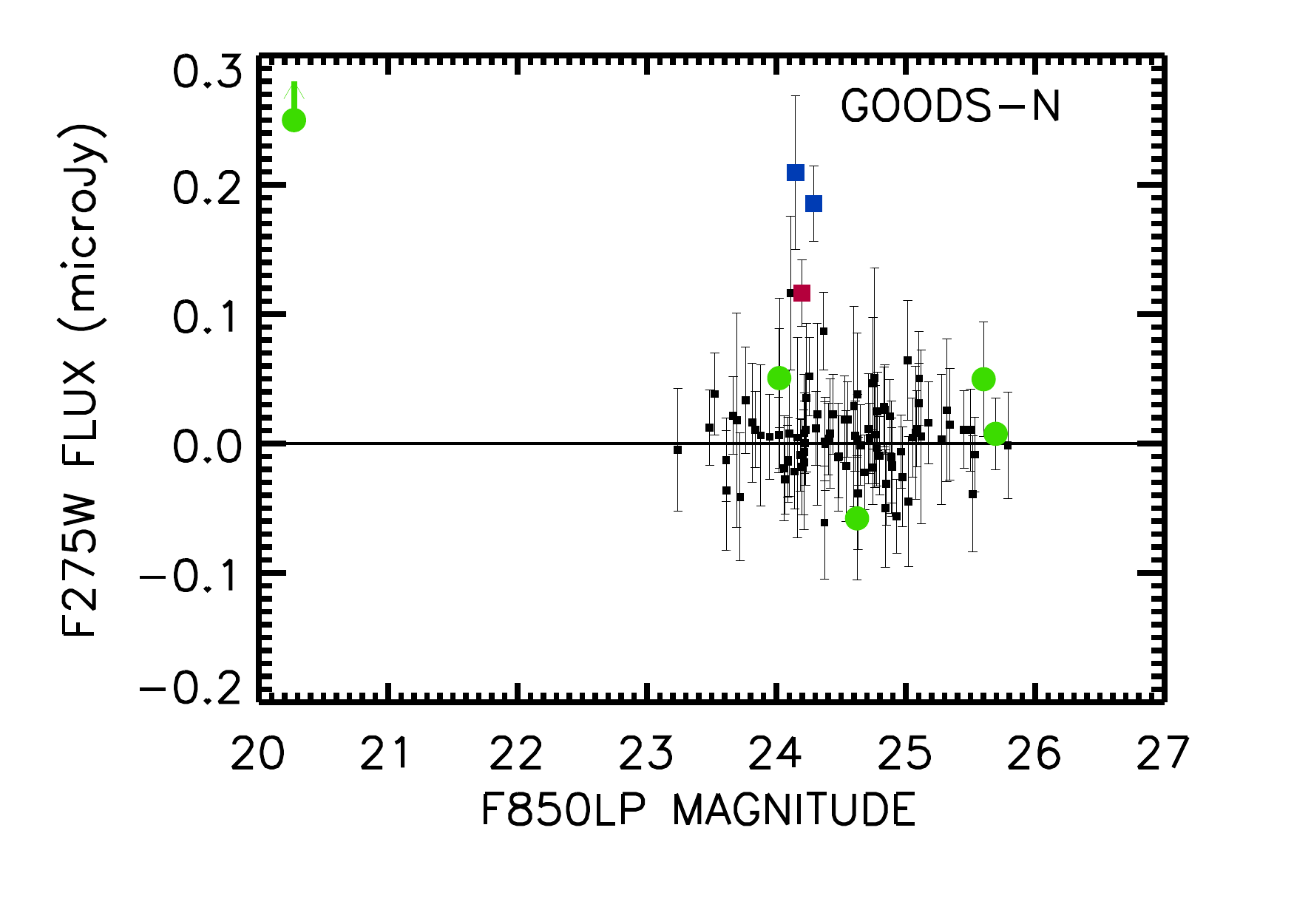}
\caption{F275W flux vs. F850LP magnitude for all $2.35 < z < 3.05$ galaxies with high-quality spectra that lie within the F275W footprint of the HDUV survey in the GOODS-S (\textit{top}) and GOODS-N (\textit{bottom}).  Sources with X-ray counterparts are marked in green, while $>3\sigma$ detections, both positive and negative, are marked in either red or blue (the latter is used to indicate sources contaminated by foreground galaxies).
Note that GN-UVC-6 is a $2.9\sigma$ source and hence not marked in red in the lower panel.
GN-UVC-1, the GOODS-N quasar shown in the lower panel, is marked in green with an upward pointing arrow, since its measured flux is much higher than the y-axis limit.}
\label{zm_flux}
\end{figure}

We select all $2.35 < z < 3.05$ galaxies with high-quality spectra that lie within the F275W footprint of the HDUV survey in both GOODS fields. For each field and band subsample, we applied the same photometric procedure that we used in our search for individual candidate LyC leakers in Section~\ref{uvsample} in order to determine the flux and error for all the sources, including those not detected individually. In Figure~\ref{zm_flux}, we show F275W flux versus F850LP magnitude for our F275W subsamples in the GOODS-S and the GOODS-N.

We next excluded sources with X-ray counterparts and sources whose spectra clearly include emission features from a line-of-sight foreground contaminant. The X-ray non-detection criterion precludes any major contributions from AGNs to our stacks. For the GOODS-S, we use the 7~Ms \textit{Chandra} Deep Field-South catalog (\citealt{luo17}; see Section~\ref{xrays}). For the GOODS-N, we use the 2 Ms \textit{Chandra} Deep Field-North catalog (\citealt{cdfn,xue16}).  We also searched for any objects with very nearby ($< 2\arcsec$) projected neighbors that might affect our aperture flux measurements, but we did not find any. Our upper redshift bound is approximately the redshift at which the F336W filter begins to probe only rest-frame wavelengths $<912$~\AA. 

We then computed the error-weighted mean flux of all the sources in each subsample and multiplied that value by the number of sources in the subsample. Because the flux errors for our sources are very similar to one another, this is quite similar to computing an unweighted sum. In the GOODS-S, this selection yielded 38 sources. In the GOODS-N, where the areal coverage in F275W is wider than in F336W, this yielded 91 sources with F275W coverage and 69 sources with coverage in both bands. In Table \ref{tab:stacktable}, we summarize the measured properties for the individual and combined GOODS fields.

In the GOODS-S, the overwhelming majority of the total contribution at F275W (0.46 $\mu$Jy of 0.51 $\mu$Jy) comes from the three candidate LyC leakers listed in Table \ref{tab:sources}. If we remove individual candidates from the stack (specifically, source 24639, which appears to have no non-ionizing emission coincident with its F275W detection), the total F275W flux drops to 0.42 $\mu$Jy, of which 0.32 $\mu$Jy is attributable to the remaining two candidates. Removing all three candidates from the stack reduces the error-weighted mean to 0.002 $\pm$ 0.006 $\mu$Jy, giving a total contribution of only 0.070 $\pm$ 0.21 $\mu$Jy.

{\setlength{\extrarowheight}{3pt}
\begin{table}[tb]
    \normalsize
    \centering
    \caption{\\\textsc{Properties of the Error-Weighted Summed Fluxes}}
    \begin{tabular}{c c c c c}
    \hline
    \hline
    Field & F275W & (SNR) & F336W & (SNR)\\
    & ($\mu$Jy) & & ($\mu$Jy) &\\
    \hline
    GOODS-S & 0.51 & (2.3) & 2.83 & (21)\\
    GOODS-N & 0.52 & (1.5) & 6.74 & (23)\\
    Both & 1.00 & (2.5) & 9.10 & (30)\\
    \hline
    \end{tabular}
\label{tab:stacktable}
\tablecomments{The area covered in the GOODS-N in F336W is
smaller than that in F275W. \\}
\end{table}
}

In order to check the background, we randomized the positions of the GOODS-S subsample and measured the magnitudes at these positions. We then processed the sample in the same way as the real sample, i.e., by eliminating low redshift sources, X-ray sources, or very nearby bright neighbors. We measured 120 random samples and found a mean background of 0.070 $\pm$ 0.023 $\mu$Jy, which is negligible compared with the measured value in the real sample. 15\% of the random samples had mean backgrounds that exceeded the total contribution of the real sample, which may suggest that we are slightly underestimating the true noise. However, it should be emphasized that we did not carefully curate the random samples in the same way that we did the real sample, and, thus, we may have a higher degree of unrecognized foreground contamination, which could result in an overly large dispersion.

Despite the greater number of sources in the GOODS-N, the total contribution at F275W is similarly small at just 0.52 $\mu$Jy.  As with the GOODS-S, a large fraction of the total contribution comes from just a small number of sources; here $\sim40$\% (0.21 $\mu$Jy) is attributable to just two sources. These are GN-UVC-6, a color-selected LyC-leaking candidate at $z = 2.439$ that we discussed in \citet{jones18} (a $2.9\sigma$ detection in F275W), and a $z = 2.98$ galaxy with an F275W magnitude of $\sim$26.21 (an $\sim3\sigma$ detection). The latter object was not discussed in \citet{jones18}, because we considered only high-redshift objects detected at $\geq 4\sigma$ significance in constructing our F275W sample, and the source also did not turn up in our color-selected sample, which probed to lower significance. In any case, this leaves $\sim$ 0.3 $\mu$Jy of F275W unaccounted for, which must come from the remaining 89 sources.

Finally, if we combine the two fields, we calculate a total contribution of 1.00 $\pm$ 0.41 $\mu$Jy at F275W, of which about 0.67 $\mu$Jy is directly attributable to the five LyC-leaking candidates discussed above (3 in the GOODS-S and 2 in the GOODS-N). Removing an individual candidate from this calculation (specifically source 24639, for the reasons mentioned in Section \ref{lyccandidates}) does not substantially change this result: the remaining 128 galaxies combined contribute 0.87 $\pm$ 0.39 $\mu$Jy, of which 0.53 is directly attributable to the remaining candidates. Given the error on the total contribution, both cases are consistent with the remainder of our sample contributing little or nothing to the overall ionizing flux output at $z \sim 2.5$.

\begin{figure}[t]
\centering
\begin{tabular}{c c}

  \includegraphics[width=0.45\linewidth]{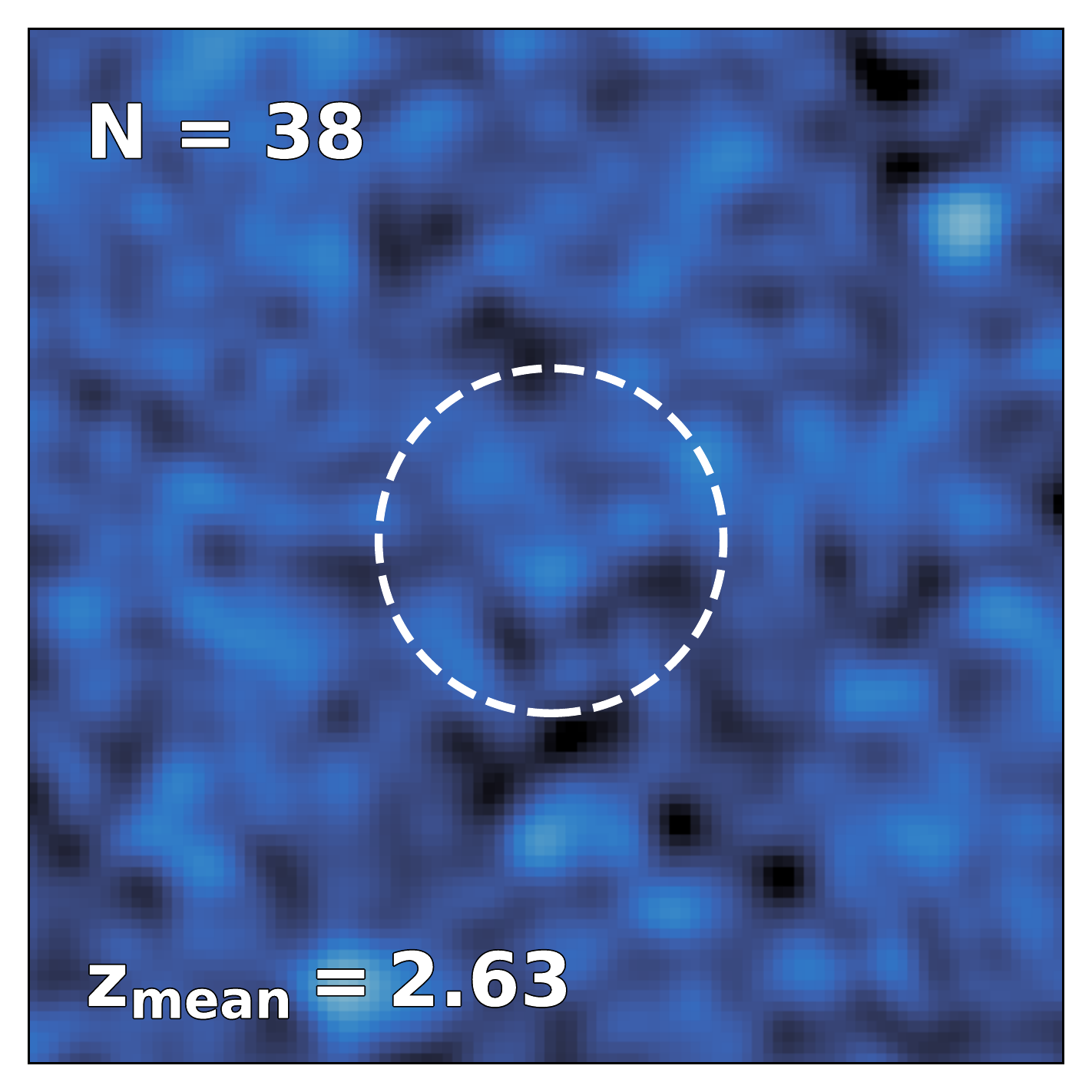} &
  \includegraphics[width=0.45\linewidth]{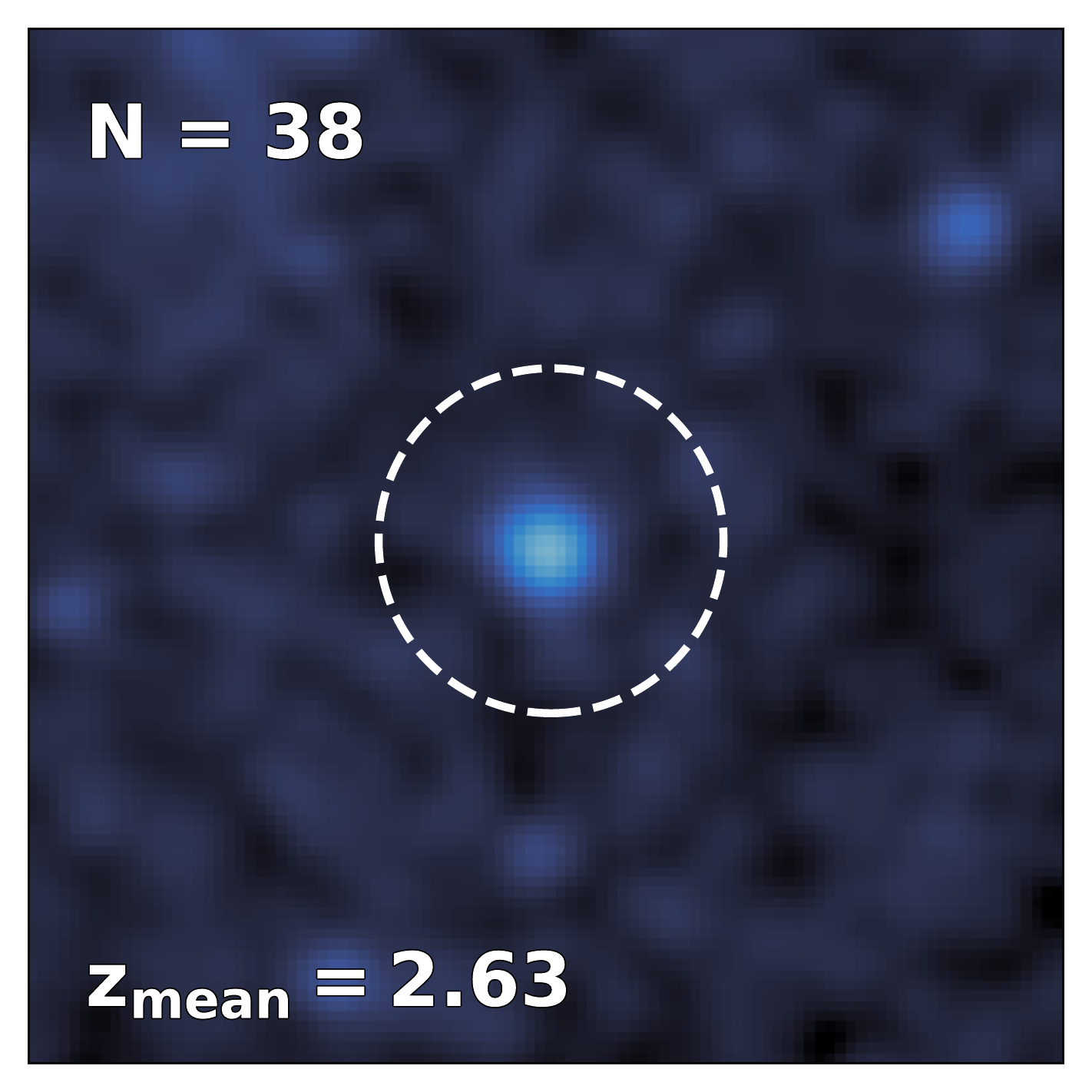}\\
  
  \includegraphics[width=0.45\linewidth]{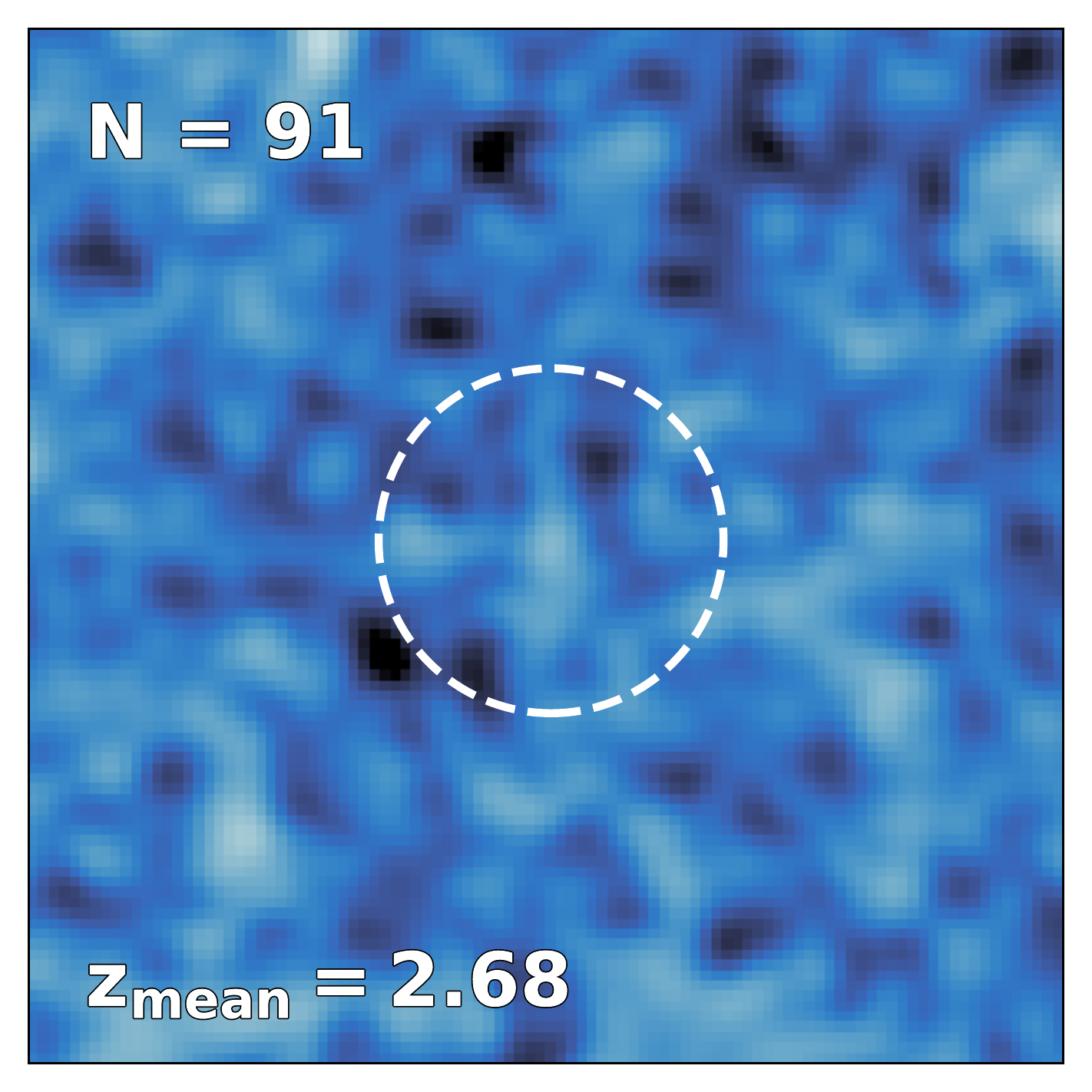} &
  \includegraphics[width=0.45\linewidth]{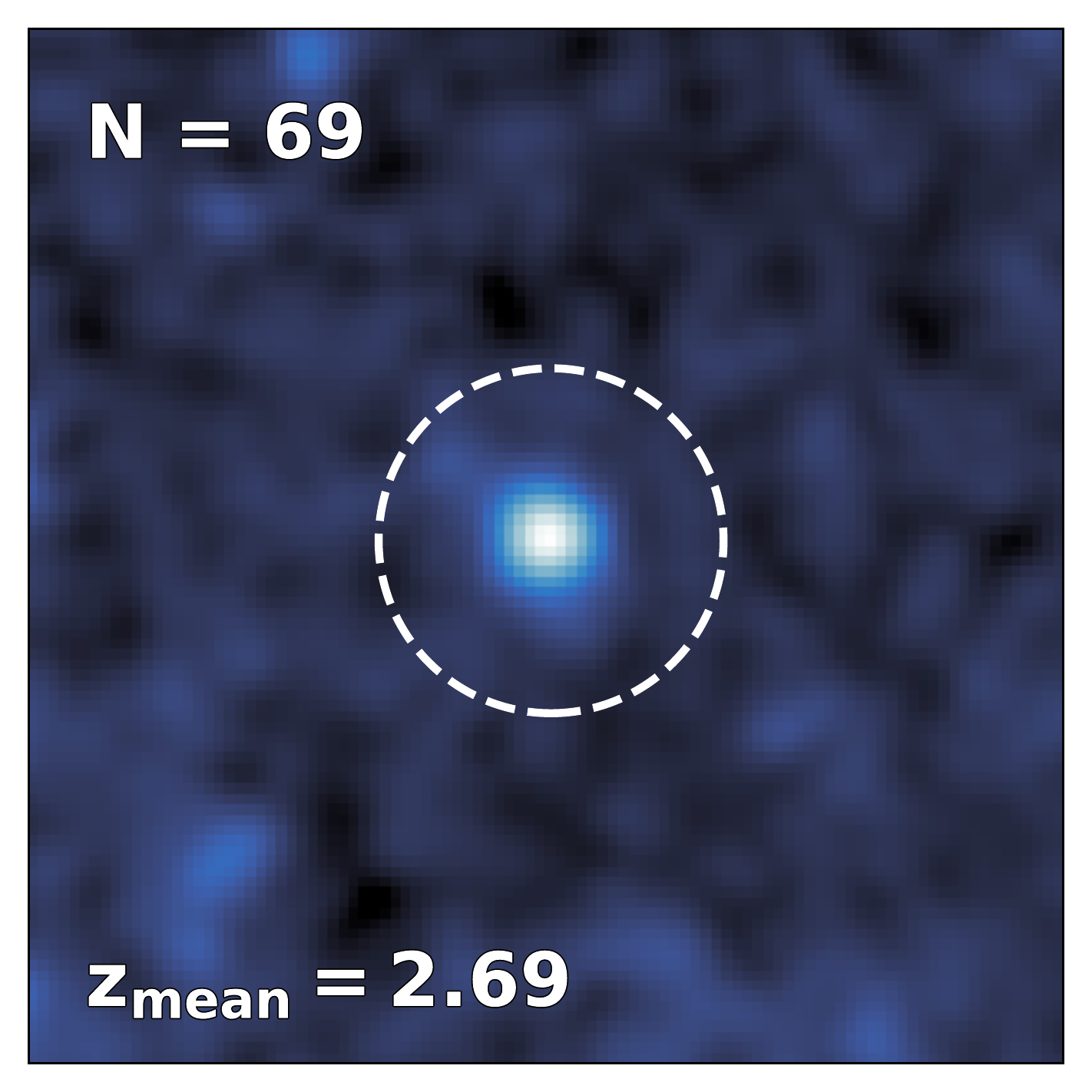}\\
  
  \includegraphics[width=0.45\linewidth]{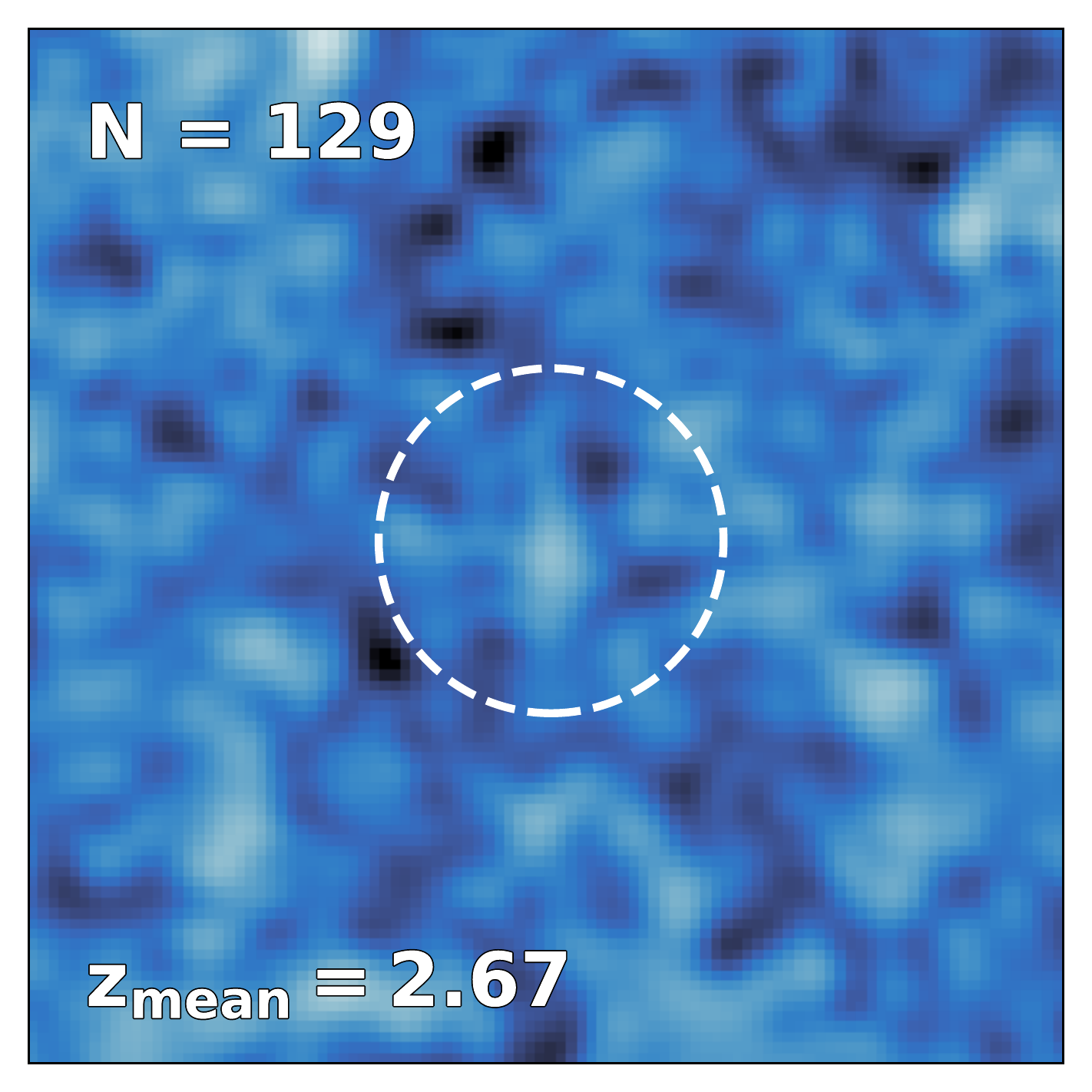} &
  \includegraphics[width=0.45\linewidth]{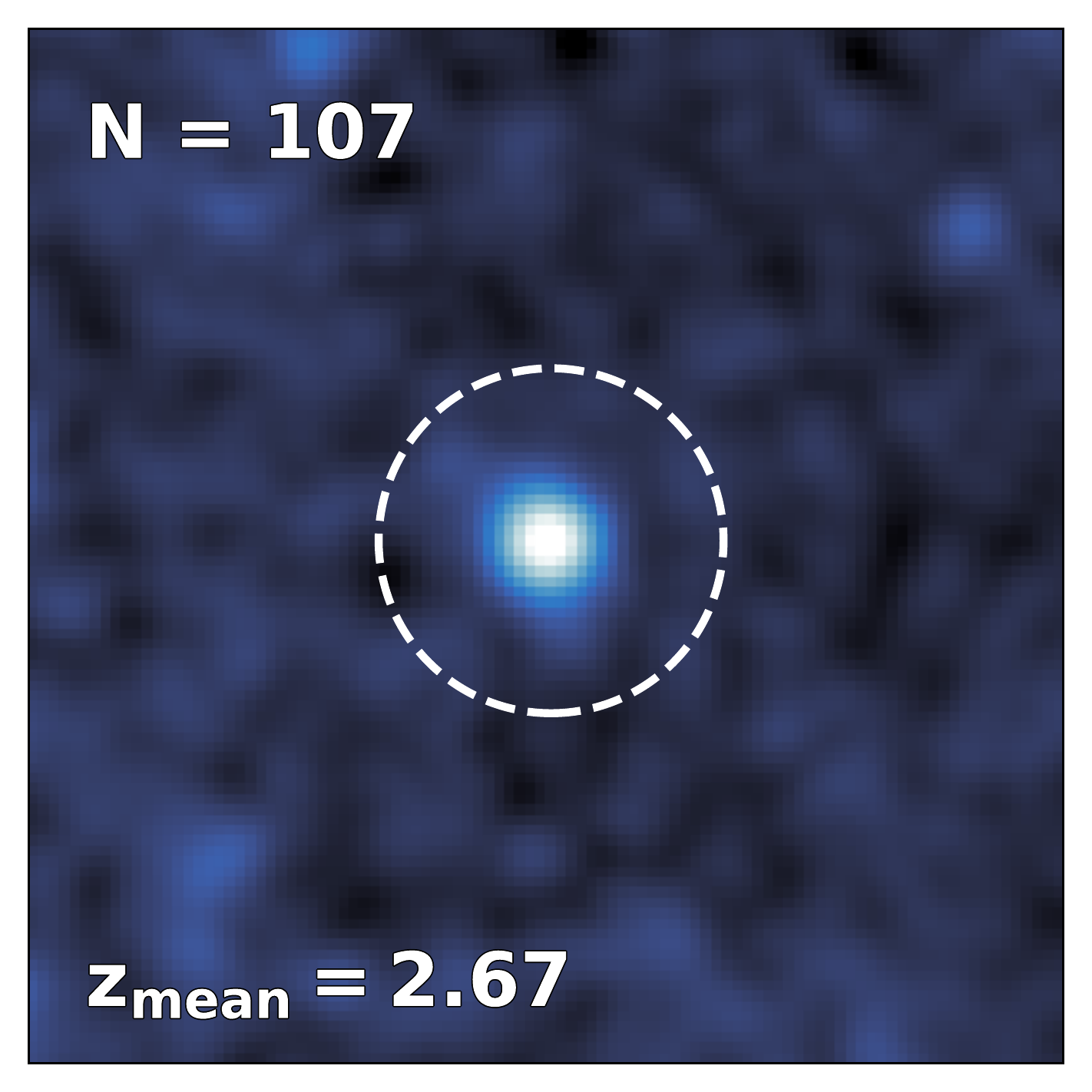}\\

\end{tabular}
\caption{
F275W (\textit{left}) and F336W (\textit{right}) stacks of spectroscopically identified sources between $2.35 < z < 3.05$ with UV coverage from the HDUV survey. Images from top to bottom show stacks from the GOODS-S, the GOODS-N, 
and both fields combined. Images are 6\arcsec~on a side and stretched to the same pixel values for each band (except for the F336W image of the combined GOODS fields, which saturates at the color scale used for the individual fields). North is up and East is to the left. The differing number of sources in the GOODS-N stacks (middle row) are due to wider areal coverage of the field in F275W than in F336W.
}
\label{totalstack_fig}
\end{figure}  

\section{An Image stacking visualization in both GOODS fields}
\label{stacking}

For visualization purposes, in Figure \ref{totalstack_fig}, we show stacked images in each band for the sample in each GOODS field, as well as for both fields combined. Bright points more than 1\arcsec~away from the image center (for example, the bright spot to the upper right in the first row of Figure \ref{totalstack_fig}) are likely foreground galaxies with significant F275W flux, though these do not affect our measurements at $z > 2.35$. Most of our candidate LyC leakers across both fields lie at $z \sim 2.4 -  2.7$, just above our redshift threshold of 2.35. At higher redshifts, it is increasingly likely that a Lyman limit system along some line-of-sight will wipe out any emergent ionizing flux, 
leading to a greater number of non-detections (see Section~\ref{flux}). 
Because of this, it is possible that including sources at the higher-redshift end of the range we probe here in the stacks could 
``dilute" some underlying average signal from sources at the lower-redshift end. 

To test this, we broke apart our stacks for the combined GOODS fields into three redshift bins, each with equal numbers of galaxies, and performed the same analysis. At F275W, we found nondetections in all three redshift bins. This suggests that the ``dilution" does not affect our ability to detect possible LyC-leaking galaxies at the depth of the HDUV F275W data. Meanwhile, the stacked detection in F336W remains fairly strong in all three redshift bins, dropping from SNR = 21.6 in the lowest-redshift bin to SNR = 5.7 in the highest, where most of the band now probes the rest-frame Lyman continuum.

In summary, we are left with (1) a handful of candidate LyC-leaking galaxies that appear to be individually detected at moderate significance (at the $\gtrsim3\sigma$ level), and (2) a general population of galaxies that are essentially invisible at LyC wavelengths, even in stacked images. This has some parallels to the results of the LACES program \citep{laces}, which used deep WFC3/UVIS imaging in F336W of the SSA22 field to search for $z \sim 3.1$ LyC leakers that are also Ly$\alpha$ emitters and Lyman break galaxies. While they successfully detect significant LyC flux in $\sim20\%$ of their targets, the majority of their sources (42 of 54) are faint in F336W (SNR $<$ 4). Stacking these non-detections yielded no net signal, with an upper limit of F336W = 31.8 (3$\sigma$ in a 0.12\arcsec aperture). This led the authors to conclude that detection of LyC emission in their sample is dichotomous, occurring either fairly strongly in individual sources or not at all. 

However, we note two important differences in methodology that distinguish this work from \citet{laces}. First, their high success rate in identifying LyC leakers (for comparison, we have a $\sim6.5\%$ success rate in the GOODS-S---that is, three candidates  out of 46 total sources with robust redshifts in the $z=2.35-3.05$ range and with UV coverage) can be attributed largely to (1) the depth of their F336W exposures compared to the typical depth of an HDUV pointing ($\sim$30 mag at 3$\sigma$ in a 0\farcs{12} diameter aperture in LACES vs. $\sim$27.5 mag at 5$\sigma$ in a 0\farcs{4} diameter aperture in HDUV), and (2) the higher throughput in F336W compared to F275W, which partially offsets the rising opacity of the IGM to LyC photons at $z > 3$. Second, their \textit{HST}/WFC3 followup was highly targeted, focusing on a set of sources that were already known to have high Ly$\alpha$ equivalent widths and large O32---both of which may correlate with LyC escape \citep{nakajima14,izo16a,izo16b,izo18,mich17,marchi18,stei18}. In contrast, the sample presented here is virtually blind to such selection effects, since we require only that a source be reasonably bright in F275W and that it have a spectroscopic redshift.

\begin{figure}[t]
\includegraphics[width=3.33in,trim={0.5cm 1cm 1cm 0},clip]{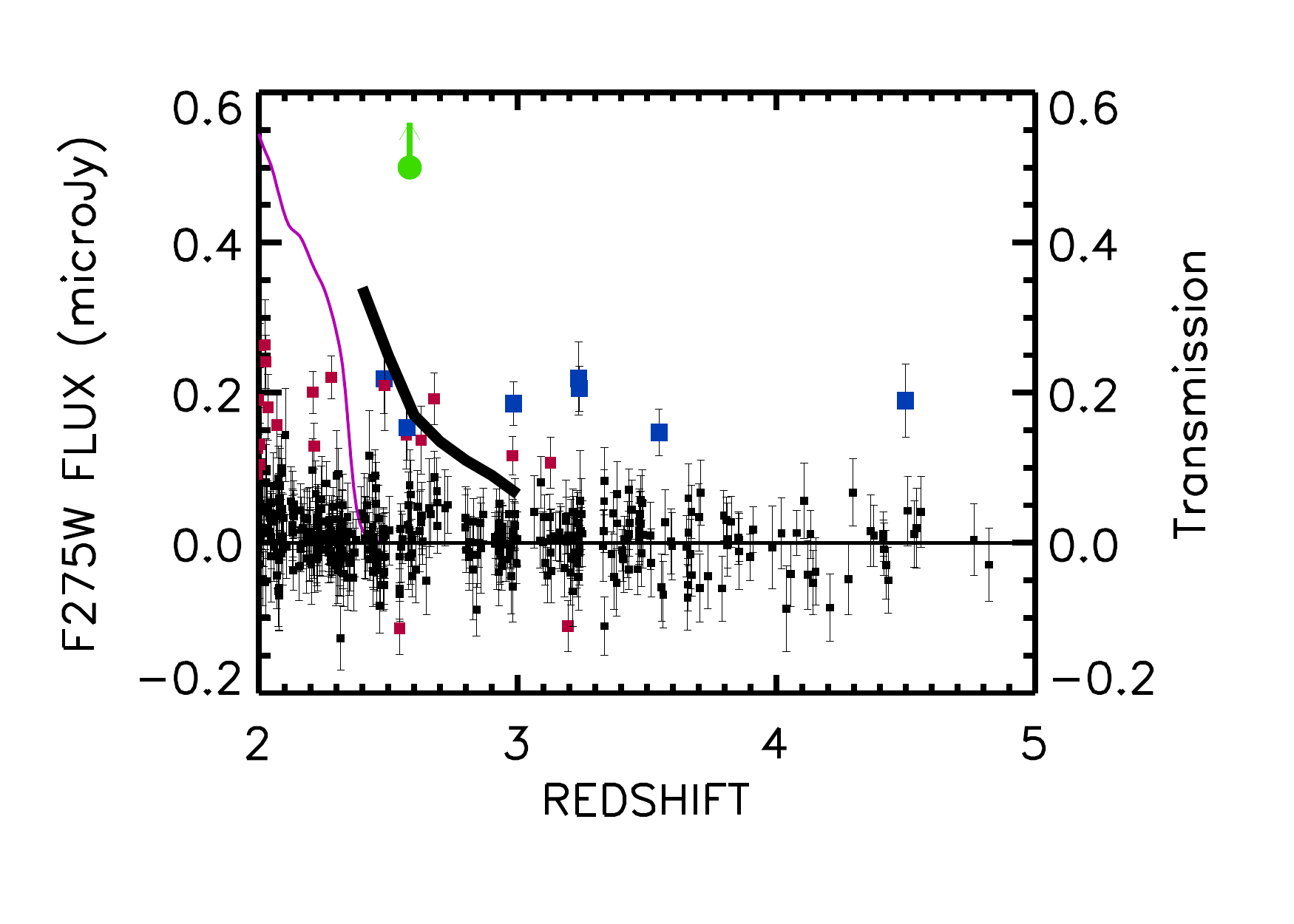}
\caption{
F275W flux vs. redshift for the 129 galaxies with high-quality spectroscopic redshifts and F275W coverage in both GOODS fields. 
The quasar in the GOODS-N (GN-UVC-1 from \citealt{jones18})
is marked in green with an upward pointing arrow, since its measured flux is much higher than the y-axis limit. The red and blue
squares are all the $>3\sigma$ detections, positive or negative, with the blue denoting
contamination by foreground galaxies. 
Note that GN-UVC-6 is a $2.9\sigma$ source and hence not marked in red.
Thus, there are five $>3\sigma$ LyC-leaking galaxy candidates in the redshift range 
$2.35 < z < 3.05$.
The thick black curve shows the transmission over this redshift range (right-hand y-axis scale), 
which drops rapidly as one moves to higher redshifts.
The purple curve shows the F275W filter transmission at the LyC edge as a function
of redshift. The y-axis units are arbitrary.
Above $z=2.35$, we will only have LyC photons within the filter.
}
\label{transmission}
\end{figure}  

\section{Contributions to the Metagalactic Ionizing Background}
\label{flux}

Determining the absolute escape fraction for each of our candidate LyC leakers is challenging, if not impossible, without reliable knowledge of each source's intrinsic spectral energy distribution (SED) and degree of reddening. Instead, we consider the extent to which our candidate sample contributes to the metagalactic ionizing background. To do so, we combine
the GOODS-N and the GOODS-S data. Excluding the spectroscopically
contaminated sources, we have a total of 129 SFGs together
with 9 AGNs based on the X-ray properties. We have identified six ofthese as candidate LyC leakers. These consist of a single quasar (the LyC-luminous source GN-UVC-1 from \citealt{jones18})
and the five SFGs described in Section~\ref{averaging}. 
GN-UVC-1 dominates the flux in the two fields with a F275W flux of 2~$\mu$Jy, roughly twice the contribution of all of the SFGs even when including the candidate LyC leakers ($1.00 \pm 0.41$ $\mu$Jy). The remaining AGNs contribute only an insignificant total of 0.096~$\mu$Jy. Because of the sparseness of the luminous AGNs, the fields are too small to make an accurate estimate of the AGN contributions, and we focus solely on the SFGs.

We converted each SFG's F275W flux density to an ionizing volume emissivity, $\epsilon_{900}$, defined as the luminosity 
density at 900 \AA~per unit frequency, divided by the comoving survey volume based on (1) the total survey area in 
F275W (131 arcmin$^{2}$) and (2) the redshift bounds $2.35 < z < 3.05$. 
We simplified our calculations by assuming that the entirety of  each SFG's F275W flux falls at the filter's effective 
wavelength of $\sim2714$~\AA, or rest-frame $\sim$750 \AA. We further assumed that the emergent SEDs of these 
galaxies are flat in frequency space (i.e., \textit{f$_{\lambda} \propto \lambda^{\beta}$} with $\beta = -2$) both blueward 
and redward of 912 \AA, albeit with different amplitudes on each side of the Lyman limit. While a shallower spectral slope 
at $\lambda_{rest} < 912$ \AA~may be more plausible, the assumption of a spectrally flat LyC allows us to use our 
measured flux densities ``as-is", without the need for scaling to 900 \AA, which would only increase our calculated 
values of $\epsilon_{900}$. This assumption thus allows us to remain as conservative as possible when estimating 
the ionizing emissivities.

Intergalactic transmission losses are substantial for the higher redshift galaxies seen in the F275W band, and 
these losses need to be taken into account in computing the ionizing emissivity. Below the LyC wavelength, 
the opacity is dominated by sources with $N(HI)$ column densities near the Lyman limit ($\log N(HI)=17.2$~cm$^{-2}$). 
Because the number of absorbing systems is small, there is a wide variation in the mean flux in the F275W filter, after
intergalactic absorption, relative to the galaxy flux at rest-frame 900~\AA\ (e.g., \citealt{in14} and references therein). 
The transmission also drops rapidly with increasing galaxy redshift as the intergalactic path length increases.

In order to compute the transmission correction, we ran Monte Carlo simulations,
 assuming the number of $N(HI)$ systems per unit $N(HI)$ is a power law with index $-1.7$,
 and the number of Lyman limit systems with $\tau > 1$ per unit $dz$ is given by the analytic 
 form $2.8\times ((1+z)/4.5)^{1.94}$ of \citet{song10}. In Figure~\ref{transmission}, we show 
 the mean flux in the F275W filter relative to the galaxy flux at rest-frame 900~\AA\ 
 versus redshift, which drops from 0.37 at $z = 2.35$ to 0.065 at $z = 3$. Because of this 
 rapid drop in the mean transmission, we have chosen to divide the sample into two redshift 
 intervals in computing the corrected emissivity. We find a logarithmic value for $\epsilon_{900}$ of
$25.0\pm0.3$ erg ~s$^{-1}$~Hz$^{-1}$~Mpc$^{-3}$ in the $z = 2.35$ -- 2.7 interval 
and $25.2\pm0.5$ erg ~s$^{-1}$~Hz$^{-1}$~Mpc$^{-3}$ in the $z = 2.7$ -- 3.05 interval. 
Removing source 24639 from our calculation in the lower-redshift interval decreases the value of log$(\epsilon_{900})$ by about 0.15 dex.

\begin{figure}[t]
\centering
\includegraphics[trim={7pt 0 8pt 0},clip,width=0.9\linewidth]{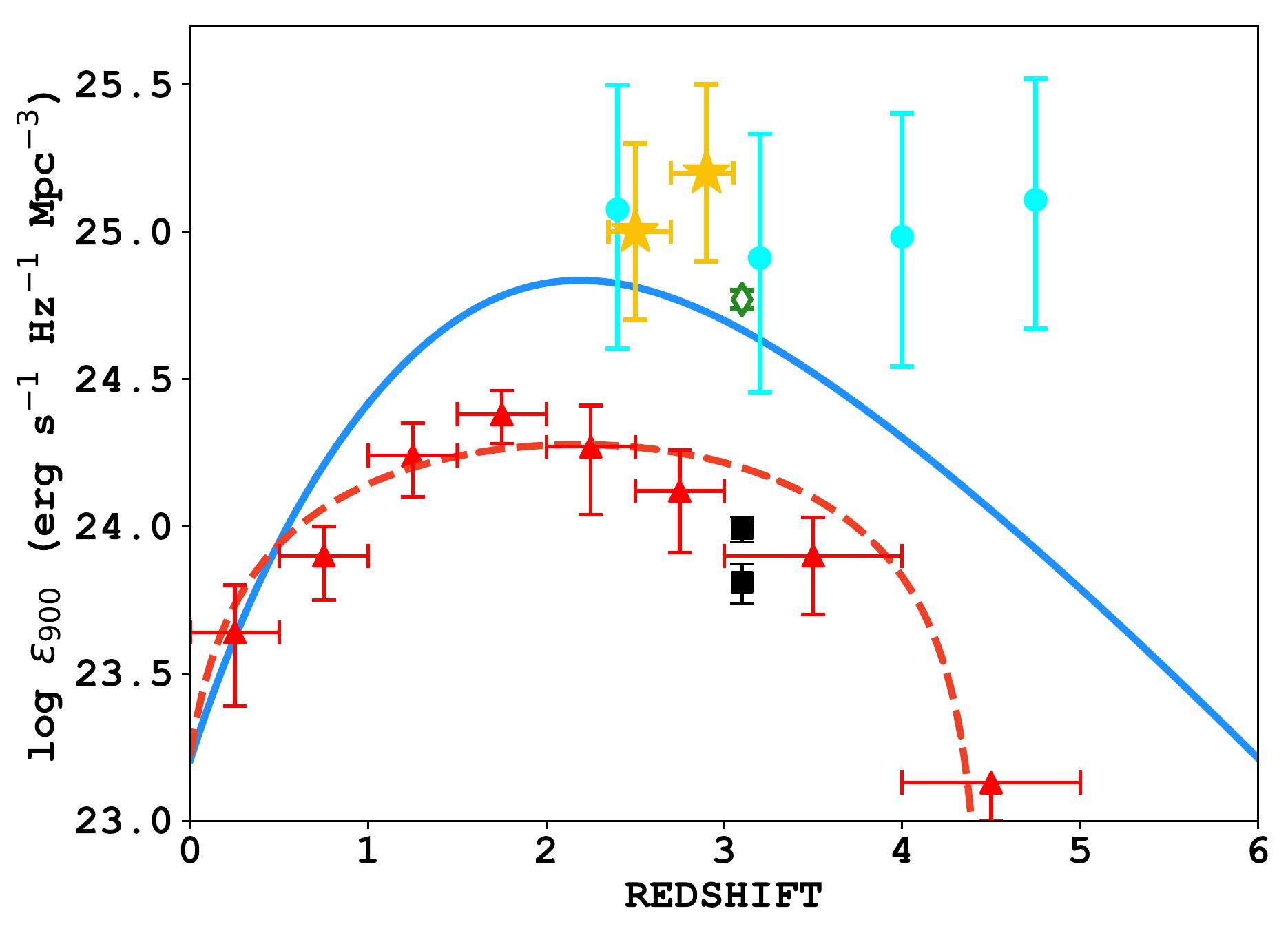} 
\caption{Ionizing volume emissivity at $z=2.35-2.7$ and $z=2.7-3.05$  estimated from our 5 LyC-leaking candidates 
together with our stacking analysis of the remaining
$z=2.35-3.05$  galaxies in the combined HDUV-GOODS fields (filled gold stars). 
Horizontal error bars on our data points indicate the redshift range. 
To compare with literature observations of the ionizing emissivity from AGN, we show data points from \citet{cbt09} and \citet{mich17} (red triangles and black squares, respectively) and best-fit models from \citet{cbt09} and \citet{ hm12} (red and blue curves, respectively). For the contribution from SFGs, we show the data point from \citet{stei18} (open green diamond). Cyan circles show the total inferred metagalactic ionizing luminosity density from \citet{beck13}.}
\label{literature}
\end{figure}  

The bulk of our ionizing flux ($67\%$) comes from the five SFGs, with the remaining galaxies contributing only a small fraction. This is consistent with the results of \citet{smith18,smith20}, who do not detect significant LyC emission in their own stacking analyses. This bimodality may be caused by favorable lines-of-sight through the ISM of the host galaxy (e.g., \citealt{cenkimm15}) and/or the IGM rather than by the intrinsic galaxy properties, with the small number of candidate LyC leakers corresponding to lines-of-sight with low IGM absorption. Even at  $z=2.35$, 14\% of the lines-of-sight have $<10$\% transmission in the F275W filter, and only 30\% of the lines-of-sight have transmission above 50\%, while at $z=2.7$, 48\% of the lines-of-sight have $<10$\% transmission in the F275W filter, and only 5\% of the lines-of-sight have transmission above 50\%. This is consistent with the redshift distribution of the candidate LyC leakers (Figure~\ref{transmission}), four of which lie in the lower-redshift interval ($z=2.35-2.7$).

In Figure \ref{literature}, we summarize the results of our calculations and put them into context using other $z\sim3$ measurements of the ionizing background from the literature. The level of ionizing volume emissivity that we estimate from the SFGs is roughly consistent with other recent estimates, such as that of \cite{stei18} at slightly higher redshifts. It is above the AGN ionizing output measured by \citet{cbt09}, \citet{hm12}, and \citet{mich17} in large samples of quasars and broad-line AGNs at similar redshifts. \citet{beck13} used Ly$\alpha$ forest observations to infer the total ionizing background from $2 < z < 5$.  At $z = 3.2$, they estimated a nominal log$(\epsilon_{900}) = 24.9$~erg~s$^{-1}$~Hz$^{-1}$~Mpc$^{-3}$, which is comparable to our estimates of $\epsilon_{900}$ for the SFG ionizing background at these redshifts. However, we emphasize that there may be further contamination in our sample by intervening sources that we have not identified. Hence, the present values should be considered as upper limits.

\section{Summary}
\label{summary}

We have presented a search for candidate LyC leakers at $z \sim 2.5-3$ in the GOODS-S field 
using a combination of deep {\em HST}/WFC3 F275W imaging data from the HDUV survey \citep{oesch18}
and extensive optical/NIR  spectroscopy. We found four candidate ionizing sources brighter than F275W $= 26$,
plus one additional source just below this cutoff with very blue rest-frame UV colors. However, two of these 
five sources are contaminated by foreground lower redshift
galaxies, as seen in their optical spectra, leaving only three possible
candidate LyC leakers.  

We performed an averaging analysis of all the sources with spectroscopic redshifts $2.35 < z < 3.05$ 
in the HDUV areas in both the GOODS-S and GOODS-N and found that the total ionizing output 
of galaxies at these redshifts is dominated by just five individual candidate LyC leakers.
These include the three GOODS-S sources described above, and two sources in the GOODS-N
(one of which was presented in \citealt{jones18}). 
Allowing for the very substantial effects of intergalactic absorption, we found that the volume ionizing
flux roughly matches that required to ionize the IGM at these redshifts (\citealt{beck13}) and is consistent 
with other recent estimates.

\acknowledgements
We thank R. Bacon and the MUSE-Deep team for providing MUSE spectra of some of our LyC-leaking candidates
ahead of publication. Some figures in this work use colormaps from the Python package \texttt{CMasher} \citep{cmasher}.
We gratefully acknowledge support from NSF grant AST-1715145, the William F. Vilas Estate, and the Kellett Mid-Career Award from the University of Wisconsin-Madison Office of the Vice Chancellor for Research and Graduate Education with funding from the Wisconsin Alumni Research Foundation (A. J. B.). Based in part on data obtained at the W. M. Keck Observatory, which is operated as a scientific partnership among the California Institute of Technology, the University of California, and NASA and was made possible by the generous financial support of the W. M. Keck Foundation. 
The authors wish to recognize and acknowledge the very significant cultural role and reverence that the summit of Mauna Kea has always had within the indigenous Hawaiian community. We are most fortunate to have the opportunity to conduct observations from this mountain.

\newcommand{\aanda}{A\&A}

\end{document}